\documentclass[aps,prd,preprint,groupedaddress]{revtex4}
\usepackage{graphicx}

\begin{document}

\preprint{OU-HEP-459}

\title{$\mbox{\boldmath $\pi N$}$ sigma-term and chiral-odd twist-3
distribution function $\mbox{\boldmath $e(x)$}$\\
of the nucleon in the chiral quark soliton model}


\author{Y.~Ohnishi and M.~Wakamatsu}
\email[]{ohnishi@kern1.nclth.osaka-u.ac.jp}
\email[]{wakamatu@phys.sci.osaka-u.ac.jp}
\affiliation{Department of Physics, Faculty of Science, \\
Osaka University, \\
Toyonaka, Osaka 560, JAPAN}



\begin{abstract}
The isosinglet combination of the chiral-odd twist-3 distribution function
$e^u(x)+e^d(x)$ of the nucleon has outstanding properties that its
first moment is proportional to the well-known $\pi N$ sigma-term and
that it contains a $\delta$-function singularity at $x=0$.
These two features are inseparably connected in that the above sum rule
would be violated, if there is no such a singularity in $e^u(x)+e^d(x)$. 
Very recently, we found that the physical origin of this 
$\delta$-function singularity can be traced back to the long-range
quark-quark correlation of scalar type, which signals the spontaneous
chiral symmetry breaking of the QCD vacuum.
The main purpose of the present paper is to give
complete theoretical predictions for the chiral-odd twist-3 distribution
function $e^a(x)$ of each flavor $a$ on the basis of the chiral quark
soliton model, without recourse to the derivative expansion type
approximation. These theoretical predictions are then compared with
the empirical information extracted from the CLAS data of the
semi-inclusive DIS processes by assuming the Collins mechanism only.
A good agreement with the CLAS data is indicative of a
sizable violation of the $\pi N$ sigma-term sum rule, or equivalently,
the existence of a $\delta$-function singularity in $e^u(x) + e^d(x)$.
\end{abstract}


\maketitle


\section{Introduction}

\ \ \ It is a widely accepted common belief now that the nonperturbative
dynamics of QCD (chiral dynamics) is an indispensable element to
understand high-energy deep inelastic scattering observables.
Undoubtedly, the reconfirmation of this natural fact strongly owes
to the two remarkable experimental discoveries in this
field \cite{EMC88}\nocite{EMC89}-\cite{NMC91}.
They are the unexpectedly small quark spin fraction 
of the nucleon revealed by the EMC measurement \cite{EMC88},\cite{EMC89}
and the light-flavor sea-quark asymmetry confirmed by the NMC
measurement \cite{NMC91}. The most successful theoretical studies of
parton distribution functions have been carried out within the
framework of the chiral quark soliton 
model (CQSM) \cite{DPPPW96}\nocite{DPPPW97}\nocite{WGR96}\nocite{WGR97}
\nocite{WK98}\nocite{WK99}\nocite{PPGWW99}\nocite{MW00}\nocite{MW01}
\nocite{MW03A}\nocite{MW03B}-\cite{DGPW99}, which is an effective model
of baryons maximally incorporating the spontaneous chiral symmetry
breaking of the QCD vacuum.
In fact, we claim that it is so far the only one effective model of
baryons which are able to explain the above two remarkable findings
simultaneously within the single theoretical framework
\cite{WY91}\nocite{WW00A}\nocite{MW92}-\cite{WW00B}.

Very recently, we became aware of another novel example in which
nonperturbative QCD dynamics play an unprecedented role in the
physics of parton distribution functions.
It concerns the possible existence of a delta-function singularity at
the Bjorken variable $x = 0$ in the chiral-odd twist-3 distribution
function $e(x)$ of the nucleon  \cite{S03},\cite{WO03}.
This distribution function itself, together with its first moment 
sum rule giving the familiar $\pi N$ sigma-term, have been known
for a long time \cite{JLS73}.
In spite of several interesting theoretical features, 
however, this distribution function has been thought of as an academic 
object of study, since, because of its chiral-odd nature, it does not 
appear in the cross section formula of inclusive DIS scatterings.
The situation changed drastically, however, since the CLAS 
Collaboration was able to get the first experimental information
on this interesting quantity through the measurement of the azimuthal
asymmetry $A_{LU}$ in the electroproduction of pions from deeply
inelastic scattering of longitudinally polarized electrons off
unpolarized protons \cite{CLAS}\nocite{herm01}-\cite{herm02}.

Some years ago, within the framework of perturbative QCD, Burkardt 
and Koike noticed that the first moment sum rule (or the $\pi N$
sigma-term sum rule) for $e(x)$ holds only when $e(x)$ has a 
$\delta$-function type singularity at the Bjorken variable 
$x = 0$ \cite{BK02}.
Unfortunately, the physical origin of this singular term is 
not very clear in this perturbative analysis. Very recently, two 
independent proofs were given to the fact that the physical origin of 
this $\delta$-function singularity can be traced back to 
the nonvanishing vacuum quark condensate which signals the spontaneous 
chiral symmetry breaking of the QCD vacuum~\cite{S03},\cite{WO03}.
An interesting question is whether we can verify  experimentally
the existence of this $\delta$-function singularity in $e(x)$.
Unfortunately, the point $x = 0$ is experimentally inaccessible.
This means that, if there really exists such a $\delta (x)$-type
singularity in $e (x)$, the experimental measurement would rather 
confirm violation of this $\pi N$ sigma-term sum rule.
Nonetheless, since $e (x)$ in the region $ x \neq 0$ can in principle
be measured, theorists are challenged to explain its behavior.

The first theoretical study of $e (x)$ was done by using the MIT bag 
model \cite{JJ92}. (See also \cite{Signal97}.)
However, this estimate based on the bag model cannot be taken
as a realistic one by the following reasons.
First, its prediction for the magnitude of the $\pi N$ sigma-term is far
from reliable. Second, more seriously, it cannot reproduce the 
$\delta$-function singularity of $e (x)$. Both these features 
(they are not actually unrelated) are easily anticipated, since the 
MIT bag model is essentially a relativistic quark model with $N_c \,(=3)$
valence quark degrees of freedom only, and the reproduction of the 
nonzero vacuum quark condensate is beyond the range of applicability
of this model. The first realistic investigation of $e(x)$ was carried out
by Efremov et al. on the basis of the chiral quark soliton model but
within the ``valence" quark only approximations \cite{EGS01},\cite{EGS02}.
More recently, the present authors and Schweitzer independently
carried out more careful analysis of the contribution of the Dirac sea
quarks on the basis of the gradient expansion type approximation and
confirmed that the isosinglet combination of $e(x)$ certainly contains
$\delta$-function type singularity \cite{S03},\cite{WO03}.
After some analysis of higher derivative terms of the expansion,
however, Schweitzer retreated to the assumption that the contribution
of the Dirac sea quarks is saturated by this $\delta(x)$ term alone.
As admitted by himself, however, whether this last assumption is justified
or not is far from trivial \cite{S03}.
To confirm it, one has to carry out an 
exact numerical calculation within the model without recourse to the
gradient expansion type approximation. Furthermore, to compare the 
predictions of the model with the experimental data of the CLAS
collaboration, one must know $e^a (x)$ of each flavor $a$.
To this end, only the knowledge 
of the isoscalar combination $e^u (x) + e^d (x)$ is not enough.
We need another independent combination, i.e. the isovector distribution 
$e^u (x) - e^d (x)$. Within the framework of the CQSM, this latter 
distribution survives at the next-to-leading order in $1 / N_c$ 
expansion and it was left untouched in \cite{S03}.

In view of the above-mentioned circumstances, we think it
important to carry out an exact model calculation within the CQSM
for both of the isoscalar and isovector combinations of the chiral-odd
twist-3 distribution function $e (x)$. 
We also think it useful to analyze the first and the second moment sum
rule for $e^u (x) + e^d (x)$ and $e^u (x) - e^u (x)$ within the CQSM 
in the light of the corresponding sum rule expected in the general
framework of perturbative QCD.
The predictions of the model for $e^u (x)$ and $e^d (x)$ (as well as 
the corresponding distributions for antiquarks) are then used as initial 
distributions given at the model energy scale around $600 \,\mbox{MeV}$
(or $Q^2 \simeq 0.30 \,\mbox{GeV}^2$), and they are evolved to higher
$Q^2$ for the sake of comparison with the phenomenological information
obtained by using the CLAS measurement.

The paper is organized as follows. In sect.2, after a brief introduction
of the basic idea of the CQSM, the theoretical expression for
$e^u (x) + e^d (x)$ and $e^u (x) - e^d (x)$ are given.
The fundamental moment sum rules for these distributions are also discussed
here in some detail. Sect. 3 is devoted to the discussion of the numerical
results. Finally, in sect.4, we summarize what we have found.

\section{$\mbox{\boldmath $e(x)$}$ in the Chiral Quark Soliton Model}


The chiral-odd twist-3 quark distribution  $e^a(x)$ of flavor $a$
inside a nucleon with 4-momentum $P$, averaged over its spin, is
defined by
\begin{equation}
 e^a(x)=
 P^+\int_{-\infty}^{\infty} \frac{dz^-}{2\pi} \,e^{ixP^+ z^-}\langle
 N|\psi_a^{\dagger}(0)\gamma^0\psi_a(z)|N \rangle
 \bigr| _{z^+=0,z_{\perp}=0}
 \label{eq:dis1} \,\, ,
\end{equation}
where $\psi_a$ are quark fields. Similarly, the corresponding
antiquark distribution is defined as
\begin{equation}
 e^{\bar{a}}(x)=
 P^+\int_{-\infty}^{\infty} \frac{dz^-}{2\pi} \,e^{ixP^+ z^-}\langle 
 N|\psi_a^{c \dagger} (0) \gamma^0 \psi_a^c (z)|N \rangle
 \bigr| _{z^+=0,z_{\perp}=0}
 \label{eq:dis2} \,\, ,
\end{equation}
with $\psi^c$ being the charge-conjugate field of $\psi$.
Here we use the standard light-cone coordinates
\begin{equation}
 z^{\pm}=\frac{z^0\pm z^3}{\sqrt{2}},\qquad P^{\pm}=
 \frac{P^0\pm P^3}{\sqrt{2}}.
 \label{eq:dis3}
\end{equation}
The variable $x$ denotes the Bjorken variable, $x=-q^2/(2P\cdot q)$,
with $q$ being the 4-momentum transfer to the nucleon.
Taking account of the charge-conjugation property of the relevant quark
bilinear operator, one can formally extend the domain of quark distribution
functions from the interval
$0\!\le\! x\! \le\! 1$ to $-1\!\le\! x\!\le\! 1$, such that
\begin{equation}
 e^{\bar{a}}(x) = e^a(-x),\quad (0\le x\le 1),
 \label{eq:disqa}
\end{equation}
which dictates that the distribution function with negative $x$ should be
interpreted as antiquark one.

Although the above definitions of the quark and antiquark distribution
functions are frame-independent, it is convenient to perform the
actual calculation in the nucleon rest frame. In this frame,
we have $P^+\!=\!M_N/\sqrt{2}$, and the distribution function is reduced
to
\begin{equation}
 e^a(x)=M_N\int_{-\infty}^{\infty}\frac{dz_0}{2\pi} \,
 e^{ixM_Nz_0}\langle N|
 \psi_a^{\dagger}(0)\gamma^0\psi_a(z)|N\rangle|_{
 z_3=-z_0 ,z_{\perp}=0}\label{eq:dis4}\, .
 \label{eq:disprest}
\end{equation}
Throughout the paper, we will confine ourselves to two flavor case of
$u$- and $d$-quarks, and neglect strangeness degrees of freedom
in the nucleon.
Consequently, we have two independent distributions, i.e. the
isosinglet distribution $e^{(T=0)}(x) \equiv e^u(x) + e^d(x)$ and the
isovector one $e^{(T=1)}(x) \equiv e^u(x) - e^d(x)$.
In the case of $e^{(T=0)}(x)$, we simply sum up (\ref{eq:dis4}) over
the flavor components. On the other hand, for $e^{(T=1)}(x)$, 
we have to sum up the representation after inserting $\tau_3$ matrix
in (\ref{eq:dis4}).

For obtaining quark distribution functions, we must generally evaluate
nucleon matrix elements of bilocal and bilinear quark operators
containing two space-time coordinates with light-cone separation. 
The startingpoint of our theoretical analysis is the following path
integral representation of the matrix elements of a bilocal and
bilinear quark operator between the nucleon states with definite
momentum
$\mbox{\boldmath $P$}$ :
\begin{eqnarray}
  \langle N (\mbox{\boldmath $P$}) \,| \,\psi^\dagger (0) \,\gamma^0 \,
  \psi(z) \,| \,N (\mbox{\boldmath $P$})\rangle
  &=& \frac{1}{Z} \,\,\int \,\,d^3 x  \,\,d^3 y \,\,
  e^{\,- \,i \mbox{\boldmath $P$} \cdot \mbox{\boldmath $x$}} \,\,
  e^{\,i \,\mbox{\boldmath $P$} \cdot \mbox{\boldmath $y$}} \,\,
  \int {\cal D} U \nonumber \\
  &&{}\times \int {\cal D}
  \psi \,\,{\cal D} \psi^\dagger \,\,
  J_N (\frac{T}{2}, \mbox{\boldmath $x$}) \,\,\psi^\dagger (0) \,
  \gamma^0 \,\psi(z) \,\,J_N^\dagger (-\frac{T}{2}, 
  \mbox{\boldmath $y$})\nonumber \\
  &&{}\times \exp \,[\,\,i \int \,d^4 x \,\,\bar{\psi} \,\,
  (\,i \! \not\!\partial \,- \,M
  U^{\gamma_5}) \,\psi \, ] \, , \ \ \ \ \
  \label{eq:cqsm3} 
\end{eqnarray}
where 
\begin{eqnarray}
  {\cal L} \ \ = \ \ \bar{\psi} \,(\,i \not\!\partial \ - \ 
  M U^{\gamma_5} (x) \,) \,\psi  ,\hspace{10mm}
  \label{eq:cqsm4}
\end{eqnarray}
with
\begin{equation}
 U^{\gamma_5} (x) = \exp [ \,i \gamma_5 \mbox{\boldmath $\tau$}
 \cdot \mbox{\boldmath $\pi$} (x) / f_\pi \,] ,
 \label{eq:ufield}
\end{equation}
being the basic lagrangian of the CQSM with two flavors. The quantity
\begin{equation}
 J_N (x) \ \ = \ \ \frac{1}{N_c !} \,\, 
 \epsilon^{\alpha_1 \cdots \alpha_{N_c}} \,\,
 \Gamma_{J J_3, T T_3}^{\{f_1 \cdots f_{N_c}\}} \,\,
 \psi_{\alpha_1 f_1} (x) \cdots \psi_{\alpha_{N_c} f_{N_c}} (x) \,\, ,
 \label{eq:intfield}
\end{equation}
is a composite operator carrying quantum numbers $J J_3, T T_3$
(spin, isospin) of the baryon, where $\alpha_i$ the color indices, while 
$\Gamma_{JJ_3,TT_3}^{\{f_1\cdots f_{N_c}\}}$ is a symmetric matrix in spin 
flavor indices $f_i$. We start with a stationary pion field configuration
of hedgehog shape :
\begin{equation}
 \mbox{\boldmath $\pi$}(x)=f_{\pi}\hat{\mbox{\boldmath $r$}}F(r).
 \label{eq:hedgehog}
\end{equation}
Next we carry out the path integral over $\mbox{\boldmath $\pi$}(x)$ in
a saddle point approximation by taking care of two zero-energy modes,
i.e. the ``translational zero-modes'' and ``rotational zero-modes''.
Under the assumption of ``slow rotation'' as compared with the intrinsic
quark motion, the answers can be obtained in a perturbative series in
$\Omega$, which can also be regarded as a $1/N_c$ expansion.
Up to the first order in the collective rotational velocity
$\Omega$, the only surviving contribution to
$e^{(T=0)}(x)$ arises at the ${\cal O}(\Omega^0)$ term of this
expansion, since the ${\cal O}(\Omega^1)$ term vanishes identically
due to the hedgehog symmetry. On the other hand,
the first nonvanishing contribution to $e^{(T=1)}(x)$ arises
at the ${\cal O}(\Omega^1)$, since the leading ${\cal O}(\Omega^0)$
contribution vanishes due to the hedgehog symmetry.
Then, between the magnitude of the above two distributions, one may
expect the following large-$N_c$ relation :
\begin{equation}
 |e^u(x)+e^d(x)| \sim N_c \,|e^u(x)-e^d(x)|\, .
 \label{eq:N_c-rel}
\end{equation}

\subsection{Isosinglet distribution \mbox{\boldmath $e^{(T=0)}(x)$}}

The isosinglet combination of the chiral-odd twist-3 unpolarized
distribution is given by
\begin{equation}
 e^{(T=0)}(x) \equiv e^u(x)+e^d(x)=M_N\int^{\infty}_{-\infty}
 \frac{dz_0}{2\pi} \,e^{ixM_Nz_0} \langle N|\bar{\psi}(0)\psi(z)|
 N \rangle |_{z_3=-z_0,z_{\perp}=0}.
 \label{eq:e0}
\end{equation}
Following the general formalism developed
in~\cite{DPPPW96},\cite{DPPPW97},\cite{WK99}, the isosinglet
distribution in the CQSM is given in the following form :
\begin{eqnarray}
 e^{(T=0)}(x) &=& -N_c M_N
 \sum_{n > 0}  \,
 \langle n \vert \gamma^0  \delta (xM_N - \hat{p}_3 - E_n) 
 \vert n \rangle 
 \label{eq:e0xunoccu}
 \\ 
 &=& \ N_c M_N 
 \sum_{n \leq 0}  \,
 \langle n \vert \gamma^0  \delta (xM_N - \hat{p}_3 - E_n) 
 \vert n \rangle ,
 \label{eq:e0xoccu}
\end{eqnarray}
where, $|n\rangle $ and $E_n$ are the eigenstates and the associated 
eigenenergies of the static Dirac Hamiltonian 
\begin{equation}
 H = -i
 \mbox{\boldmath $\alpha\cdot\nabla $} + \beta M e^{i\gamma_5
 \mbox{\boldmath $\tau\!\cdot\!\hat{r}$}F(r)} ,
 \label{eq:Dirac.hamil}
\end{equation}
with the hedgehog background. Here, the summation $\sum_{n\le 0}$ in
(\ref{eq:e0xoccu}) is meant to be taken over the valence-quark orbital
(it is the lowest energy eigenstate that emerges from the
positive-energy Dirac continuum) plus
all the negative-energy Dirac-sea orbitals.
On the other hand, the summation $\sum_{n > 0}$ in (\ref{eq:e0xunoccu})
is meant to be taken over all the positive-energy Dirac continuum
excluding the discrete valence orbital.
We recall that the CQSM is defined with some appropriate regularization.
In fact, without regularization, $e^{(T=0)}(x)$ is quadratically divergent,
and no practical meaning can be given to either
of~(\ref{eq:e0xunoccu}) and (\ref{eq:e0xoccu}).
The ideal regularization scheme for our purpose is the Pauli-Villars
subtraction scheme, since it preserves several fundamental
conservation laws of field theory~\cite{DPPPW96},\cite{DPPPW97}.
Furthermore, it is also expected to preserve the equivalence of
the two ways of computing the quantity in question, by using
(\ref{eq:e0xunoccu}) and (\ref{eq:e0xoccu}).
In the present study, we use the double subtraction Pauli-Villars scheme
as introduced in~\cite{KWW99}, since $e^{(T=0)}(x)$
divergees like the vacuum quark condensate.
In this scheme the distribution $e^{(T=0)}(x)$ is replaced with a
regularized one defined as 
\begin{equation}
 e^{(T=0)}(x) \ \equiv \ 
 e^{(T=0)}(x)^M - \sum_{i=1}^2 c_i\left( \frac{\Lambda_i}{M}\right)
 e^{(T=0)}(x)^{\Lambda_i} .
 \label{eq:pv}
\end{equation}
Here $e(x)^{\Lambda_i}$ is obtained from $e(x)^M$  by replacing the mass
parameter $M$ by $\Lambda_i$. It was shown in~\cite{KWW99}
that, if the parameters $c_1, c_2, \Lambda_1$, and 
$\Lambda_2$ are chosen to satisfy the two conditions :
\begin{eqnarray}
 1 - \sum_{i = 1}^2 \,c_i \,\left(\frac{\Lambda_i}{M} \right)^2 &=& 0 , 
 \label{eq:cond1} \\
 1 - \sum_{i = 1}^2 \,c_i \,\left(\frac{\Lambda_i}{M} \right)^4 &=& 0 , 
 \label{eq:cond2}
\end{eqnarray}
the quadratic as well as the logarithmic divergences in the vacuum quark
condensate are completely eliminated.

Actually, we are interested in the nucleon observables measured in
reference to the physical vacuum, so that $e^{(T=0)}(x)$ should be
replaced by
\begin{equation}
 e^{(T=0)}(x)\rightarrow e^{(T=0)}(x)\equiv 
 e^{(T=0)}_U(x)-e^{(T=0)}_{U=1}(x).
 \label{eq:exregl}
\end{equation}
Here the vacuum subtraction term $e^{(T=0)}_{U=1}(x)$ is obtained from
$e^{(T=0)}_U(x)$ by setting $U=1$ or $F(r)=0$, and by excluding the sum
over the discrete valence level.
We point out that, due to the energy-momentum conservation
embedded in the factor $\delta(x M_N-\hat{p}_3-E_n)$, the vacuum subtraction
terms are required only for $x<0$ in the occupied form (\ref{eq:e0xoccu}),
and for $x>0$ in the non-occupied form (\ref{eq:e0xunoccu}).
This means that the vacuum subtraction terms need not be considered when
$e^{(T=0)}(x)$ is evaluated in the following way, i.e. if it is
evaluated by using the occupied form for $x\!>\!0$, while by
using the nonoccupied form for $x\!<\!0$.

\subsubsection{momentum sum rules of $e^{(T=0)}(x)$}

The most important information of the distribution functions are
generally contained in their first few moments of lowest orders.
This is also the case for the distribution $e^{(T=0)}(x)$.
In a recent paper, Efremov and Schweitzer reviewed some of the
important sum rules for the chiral-odd twist-3 distribution functions
in an enlightening way \cite{ES03}.
Their argument starts with the general definition
of the distribution with flavor $a$ as
\begin{equation}
 e^a(x)=\frac{1}{2M_N}\int\frac{d \lambda}{2\pi} \,e^{i\lambda x}
 \langle N|\bar{\psi}_a(0)[0,\lambda n]\psi_a(\lambda n)|N\rangle , 
 \label{eq:link}
\end{equation}
where $[0,\lambda n]$ denotes the gauge link. By using an operator
identity following from the QCD equation of motion, $e^a(x)$ is shown
to be decomposed in a gauge invariant way into the following three
pieces :
\begin{equation}
 e^a(x) = e^a_{sing}(x) + e^a_{tw3}(x) + e^a_{mass}(x).
 \label{eq:exdecomp}
\end{equation}
Here, $e^a_{sing}(x)$ denotes a singular term given by
\begin{equation}
 e^a_{sing}(x) = \delta(x) \,\langle N | \bar{\psi}^a \psi^a |
 N \rangle .
 \label{eq:exsing}
\end{equation}
On the other hand, $e^a_{tw3}(x)$ is a genuine twist-3 part of $e^a(x)$
that contains information on quark-gluon-quark correlations.
Finally, $e^a_{mass}(x)$ denotes the term arising from the nonzero current
quark mass. It is a somewhat peculiar function defined through its
Mellin moments as \cite{BBKT96}\nocite{BM97}\nocite{KN97}-\cite{KT99}
\begin{equation}
 \int_{-1}^1 \,x^{n-1} \,e^a_{mass}(x) \,dx = \delta_{n>1} \cdot
 \frac{m_0}{M_N} \,\int_{-1}^1 \,x^{n-2} \,f^a_1(x) \,dx ,
 \label{eq:exmassmom}
\end{equation}
with $f^a_1(x)$ being the twist-2 unpolarized distribution with flavor
$a$. The presence of the factor $\delta_{n>1}$ here dictates that the
first moment of $e^a_{mass}(x)$ vanishes,
\begin{equation}
 \int_{-1}^1 \,e^a_{mass}(x) \,dx = 0 .
 \label{eq:exmass1stmom}
\end{equation}
It is also known  \cite{BBKT96}\nocite{BM97}\nocite{KN97}-\cite{KT99}
that the first two basic Mellin moments of $e^a_{t w 3} (x)$
vanish, i.e.
\begin{equation}
 \int_{-1}^1 x^{n - 1} e^a_{t w 3} (x) d x = 0 \ \ \ \mbox{for} \ \ 
 n = 1,2 .
 \label{eq:extw3mom}
\end{equation}
Putting the above-mentioned properties altogether, the first moment sum 
rule for the isoscalar combination of $e^a (x)$, i.e. $e^{(T = 0)} (x)$ 
takes the form. 
\begin{equation}
 \int_{-1}^1 e^{(T = 0)} (x) d x = \frac{\sum_{\pi N}}{m_0} ,
 \label{eq:ex1stmom}
\end{equation}
which is nothing but the $\pi N$ sigma-term sum rule. Note that this sum
rule is saturated by the first term of (\ref{eq:exdecomp}) alone.
On the other hand, the second Mellin moment of $e^{(T = 0)} (x)$ is 
given by
\begin{equation}
 \int_{-1}^1 x e^{(T = 0)} (x) d x = \frac{m_0}{M_N} \cdot N_c ,
 \label{eq:ex2ndmom}
\end{equation}
where $N_c$ is the number of color, which coincides with the number of
quarks contained in a baryon-number-one system, i.e. $N_c = 3$.
We point out that this second Mellin moment of $e^{(T = 0)}$ vanishes
in the chiral limit of $m_0 = 0$.

Next, we turn to the discussion of the moment sum rule in the CQSM.
Integrating (\ref{eq:e0xoccu}) over $x$, the first moment of
$e^{(T = 0)} (x)$ is given as
\begin{equation}
 \int_{-1}^1 e^{(T = 0)} (x) d x = N_c \sum_{n \leq 0}
 \langle n | \gamma^0 | n \rangle .
 \label{eq:ex1stmom-cqsm1}
\end{equation}
Since the r.h.s. of this equation is nothing but the scalar charge 
$\bar{\sigma}$ of the nucleon within the CQSM, the sigma-term sum rule 
immediately follows
\begin{equation}
 \int_{-1}^1 e^{(T = 0)} (x) d x = \bar{\sigma} = 
 \frac{\sum_{\pi N}}{m_0} .
 \label{eq:ex1stmom-cqsm2}
\end{equation}
The way of this sum rule being satisfied is far more delicate in the CQSM
than in the above QCD-motivated analysis. As shown by our previous study,
although the model certainly predicts the $\delta (x)$-type singularity in
$e^{(T = 0)} (x)$, this term alone does not saturate the $\pi N$
sigma-term sum rule. The model also predicts nontrivial structure of
$e^{(T = 0)} (x)$ at $x \neq 0$, which may contribute to the first
moment sum rule. We shall discuss this point in more detail
in the next section.

Turning to the second moment, it is easy to show from (\ref{eq:e0xoccu})
that
\begin{equation}
 \int_{-1}^1 x e^{(T = 0)} (x) d x = \frac{N_c}{M_N}
 \sum_{n \leq 0} \langle n | \gamma^0 (\hat{p}_3 + E_n ) | n \rangle .
 \label{eq:ex2ndmom-cqsm1}
\end{equation}
Owing to the hedgehog symmetry of the soliton, the term containing 
$\gamma^0 \hat{p}_3$ vanishes, and we are left with
\begin{equation}
 \int_{-1}^1 x e^{(T = 0)} (x) d x = \frac{N_c}{M_N}
 \sum_{n \leq 0} E_n \langle n | \gamma^0 | n \rangle .
 \label{eq:ex2ndmom-cqsm2}
\end{equation}
Following \cite{S03}, it is convenient to rewrite the r.h.s. of the
above equation in the following manner.
First, notice the identity
\begin{equation}
 E_n \langle n | \gamma^0 | n \rangle
 = \frac{1}{2} \langle n | \{ \hat{H}, \gamma^0 \} | n \rangle
 = m_0 + M \langle n | \frac{1}{2} (U + U^{\dagger}) | n \rangle .
 \label{eq:anticom}
\end{equation}
Here, we have tentatively restored the current quark mass term in the 
model Hamiltonian $H$, just for the sake of explanation here only, i.e.
we have used here
\begin{equation}
 H = -i \mbox{\boldmath $\alpha$} \cdot \nabla 
 + \beta M e^{i \gamma_5 \mbox{\boldmath $\tau$} \cdot 
 \hat{\mbox{\boldmath $r$}} F(r)} 
 + m_0 .
 \label{eq:dirac.hamil.m}
\end{equation}
Then, the second moment sum rule in the CQSM takes the following form :
\begin{equation}
 \int_{-1}^1 x e^{(T = 0)} (x) d x = \frac{N_c}{M_N} \,(m_0 + \beta M) ,
 \label{eq:ex2ndmom-cqsm3}
\end{equation}
with
\begin{equation}
 \beta \equiv \sum_{n \leq 0} \langle n | \frac{1}{2} (U + U^{\dagger})
 | n \rangle .
 \label{eq:beta}
\end{equation}
It is clear now that r.h.s. of this sum rule does not vanish even in the
chiral limit of $m_0 = 0$, in contrary to the sum rule derived from the
QCD-equation-of-motion method.
We shall return to this point in the next section.

\subsection{Isovector distribution \mbox{\boldmath $e^{(T=1)}(x)$}}

The isovector distribution is defined by
\begin{equation}
 e^{(T=1)}(x) \equiv e^u(x)-e^d(x)=M_N\int^{\infty}_{-\infty}
 \frac{dz_0}{2\pi} \,e^{ixM_Nz_0} \langle N|\bar{\psi}(0)\tau_3\psi(z)|
 N \rangle |_{z_3=-z_0,z_{\perp}=0} .
 \label{eq:e1}
\end{equation}
Within the framework of the CQSM,  $e^{(T=1)}(x)$ survives
only in the next-to-leading order in the collective angular velocity
$\Omega$. Following the formalism derived in~\cite{WK99},\cite{PPGWW99},
the final answer is written in the form :
\begin{eqnarray}
 e^{(T=1)}(x)&=& -\langle 
 2T_3\rangle_p M_N \frac{N_c}{2I}\frac13\sum_{a=1}^3
   \sum_{m = all, n > 0}  
 \langle n | \tau_a | m \rangle
 \langle m | \tau_a  \gamma^0  
 \left( \frac{\delta_n}{E_m - E_n} - \frac{1}{2M_N} \,
 \delta_n^\prime \right)
 | n \rangle \nonumber\\
 &=&\langle 
 2T_3\rangle_p M_N \frac{N_c}{2I}\frac13\sum_{a=1}^3
 \sum_{m = all, n \leq 0}
 \langle n | \tau_a | m \rangle
 \langle m | \tau_a  \gamma^0  
 \left( \frac{\delta_n}{E_m - E_n} - \frac{1}{2M_N} \,
 \delta_n^\prime \right)
 | n \rangle , \ \ 
 \label{eq:dis31}
\end{eqnarray}
with $\delta_n\equiv \delta(x M_N-E_n-p^3)$ and 
$\delta'_n=\frac{\partial}{\partial x}\delta(x M_N-E_n-p^3)$.
Here $I$ in the r.h.s. of (\ref{eq:dis31}) is the moment of inertia
of the soliton, given by
\begin{equation}
 I = \frac{N_c}{6}\sum_{a=1}^3 \sum_{m > 0} \sum_{n \le 0}
 \frac{\langle n |\tau_a |m \rangle \langle m |\tau_a | 
 n \rangle}{E_m-E_n}.
 \label{eq:momentinertia}
\end{equation}
In (\ref{eq:dis31}), $\langle O\rangle_p$
should be understood as an abbreviated notation of the matrix element
of a collective operator $O$ between the (spin-up) proton state, i.e.
\begin{eqnarray}
 {\langle O \rangle}_p &\equiv& \int
 \Psi_{T = T_3 = 1/2 ; J = J_3 = 1/2}^{*} [\xi_A] \,O [\xi_A] \,
 \Psi_{T = T_3 = 1/2 ; J = J_3 = 1/2} [\xi_A] \,d \xi_A \nonumber \\
 &=& \ \ \langle p,S_3=1/2|O|p,S_3=1/2\rangle .
 \label{eq:matrcollect}
\end{eqnarray}
In the present case, we have ${\langle 2T_3 \rangle}_p = 1$.

We immediately notice that the above expressions are not suitable for
the actual numerical calculation. Here, we shall proceed as in the
previous studies~\cite{WK99},\cite{PPGWW99}. First, note that the term
containing the $x$-derivative of the $\delta$-function
in (\ref{eq:dis31}) can be rewritten as
\begin{equation}
 e_2(x)=-
 \frac{d}{dx} \frac{N_c}{4I}\frac13\sum_a
 \sum_{m = all, n \leq 0}  
 \langle n | \tau_a | m \rangle
 \langle m | \tau_a  \gamma^0 
 \delta_n 
 | n \rangle  = 
 \frac{1}{4IM_N}\frac{d}{dx}\,e^{(T=0)}(x).
 \label{eq:de0}
\end{equation}
Here, we have made of the completeness of the eigenstates
$| n \rangle$ of the static Dirac Hamiltonian $H$.
(We recall that $e_2(x)$ term originates from the non-locality in time
of the operator $\bar{\psi}(0)\tau^a\psi(z)$ in (\ref{eq:e1}).)
It should be recognized that the $x$-derivative of the isosinglet
distribution $e^{(T=0)}(x)$ appears in the right-hand side.
Since we already know that the isosinglet distribution $e^{(T=0)}(x)$
has the $\delta(x)$ type singularity connected with the nonvanishing
vacuum expectation, it then follows that $e_2(x)$ has the
derivative-of-$\delta(x)$ type singularity.
However, it is unlikely that the net isovector distribution
$e^{(T=1)}(x)$ has such a singularity, because the QCD vacuum should
not violate isospin symmetry so that vacuum quark condensate of
isovector type must simply vanish.
This apparent discrepancy can be resolved as follows.
We first divide the double sum of (\ref{eq:dis31}) into the sum over
terms with $E_m=E_n$ and with $E_m\ne E_n$.
The point is that the sum with $E_m=E_n$ in $e_1(x)$ can be rewritten
in a similar form as the corresponding term in $e_2(x)$,
\begin{eqnarray}
 e_1(x)&=&
 M_N \frac{N_c}{2I}\frac13\sum_a
 \sum_{{m = all, n \leq 0 \atop (E_m \ne E_n)}}\frac{1}{E_m-E_n}  
 \langle n | \tau_a | m \rangle
 \langle m | \tau_a  \gamma^0  
 \delta_n 
 | n \rangle
 \nonumber\\
 &&+
 \frac{d}{dx}\frac{N_c}{4I}\frac13\sum_a
 \sum_{{m\leq 0,n\leq0\atop (E_m=E_n)}}
 \langle n | \tau_a | m \rangle
 \langle m | \tau_a  \gamma^0  
 \delta_n
 | n\rangle ,
 \label{eq:dble1}\\
 e_2(x)&=&- 
 \frac{d}{dx}\frac{N_c}{4I}\frac13\sum_a
 \sum_{{m = all,n\leq0\atop (E_m \ne E_n)}}
 \langle n | \tau_a | m \rangle
 \langle m | \tau_a  \gamma^0  
 \delta_n
 | n\rangle
 \nonumber\\
 &&- 
 \frac{d}{dx}\frac{N_c}{4I}\frac13\sum_a
 \sum_{{m\leq 0,n\leq0\atop (E_m=E_n)}}
 \langle n | \tau_a | m \rangle
 \langle m | \tau_a  \gamma^0  
 \delta_n
 | n\rangle .
 \label{eq:dblsum}
\end{eqnarray}
Now, just as argued in~\cite{PPGWW99},\cite{WK99}, $E_m = E_n$
contribution in the double sums in $e_1(x)$ and $e_2(x)$ precisely
cancel each other.
After regrouping the terms in such a way that this cancellation occurs at
the level of analytical expressions, the ${\cal O}(\Omega^1)$
contribution to the distribution function $e^{(T=1)}(x)=e^u(x)-e^d(x)$
can finally be written in the following form:
\begin{equation}
 e^{(T=1)}(x)=
 M_N \frac{N_c}{2I}\frac13\sum_a
 \sum_{{m = all, n \leq 0\atop(E_m \ne E_n)}}  
 \langle n | \tau_a | m \rangle
 \langle m | \tau_a  \gamma^0  
 \left( \frac{\delta_n}{E_m - E_n} - \frac{1}{2M_N} \,
 \delta_n^\prime \right)
 | n \rangle .
 \label{eq:nosing}
\end{equation}
The fact is that, in the double sum of (\ref{eq:dblsum}),
the singularity connected with the non-zero vacuum quark condensate
comes only from $E_m = E_n$ contribution, i.e. the second term of
(\ref{eq:dblsum}). As mentioned above, after the $E_m = E_n$
contributions in $e_1(x)$ and $e_2(x)$ are canceled, these singularities
disappear in (\ref{eq:nosing}). The final theoretical formula
(\ref{eq:nosing}) is therefore free from any
singularity which contradicts the symmetries of the QCD vacuum, and
it provides us with a sound basis for numerical calculation.

\subsubsection{first moment sum rule of $e^{(T=1)}(x)$}
\label{sum-rule-e1}

 Here we discuss the first moment sum rule of the isovector
distribution. Integrating (\ref{eq:e1}) over $x$, we obtain
\begin{equation}
 \int_{-1}^1 e^{(T=1)}(x) dx = 
 \int_{-1}^1 (e^u(x)-e^d(x))dx=\langle N|\bar{\psi}
 \tau_3\psi|N\rangle \, .
 \label{eq:dis12}
\end{equation}
(Here, $\bar{\psi} \tau_3\psi$ should be taken as an
abbreviated notation of $\int \bar{\psi}(\mbox{\boldmath $y$}) 
\tau_3\psi (\mbox{\boldmath $y$}) \,d^3 \mbox{\boldmath $y$}$,
which gives the isovector scalar charge operator.)
An interesting observation is that the first moment of $e^{( T=1)}(x)$ is 
related to the non-electromagnetic mass difference of neutron and proton.
In fact, the nonelectromagnetic neutron-proton mass difference is
thought to be generated by the isospin breaking term in the QCD
Hamiltonian : 
\begin{equation}
 \Delta H=\frac{m_u-m_d}{2} \,(\bar{\psi}_u\psi_u-\bar{\psi}_d\psi_d)\, .
 \label{eq:chiralsb}
\end{equation}
Because of the smallness of all the masses $m_u,m_d,m_d-m_u$ compared
with the typical energy-scale of hadron physics, we can treat $\Delta H$
as the first order perturbation, thereby being led to the following
formula for the non-electromagnetic mass difference between neutron
and proton :
\begin{eqnarray}
 (M_n-M_p)_{QCD}&=&\langle n |\Delta H|n\rangle - 
 \langle p|\Delta H |p\rangle
 \nonumber\\
 &=& (m_d-m_u)\langle p|\bar{\psi}_u\psi_u-\bar{\psi}_d 
 \psi_d|p\rangle ,
 \label{eq:npmdif}
\end{eqnarray}
where use has been made of the isospin symmetry for the unperturbative
state $|p\rangle ,|n\rangle $ (i.e., the invariance under the
interchanges $p\leftrightarrow n$ and $u\leftrightarrow d$).
Empirically, the neutron-proton mass difference of QCD origin can be
estimated from the observed mass difference by taking account of
the correction due to the electromagnetic interactions :
\begin{equation}
 (M_n-M_p)_{QCD}=(M_n-M_p)_{exp}-(M_n-M_p)_{e.m.}\, .
 \label{eq:npmdifcor}
\end{equation}
Using the values $(M_n-M_p)_{exp}\simeq 1.29 \,\mbox{MeV},\,
(M_n-M_p)_{e.m.}\simeq (-0.76\pm 0.30)\mbox{MeV}$~\cite{GL82},
we obtain 
\begin{equation}
 (M_n-M_p)_{QCD}\simeq (2.05\pm 0.30) \,\mbox{MeV}\, ,
 \label{eq:npmdifqcd}
\end{equation}
To extract the first moment of $e^{(T=1)}(x)$ empirically, we
need to know the value of $m_d - m_u$. By using
$m_d - m_u \simeq 5 \,\mbox{MeV}$, as an order of magnitude
estimate, we obtain
\begin{eqnarray}
 \int_{-1}^1 e^{(T=1)}(x) \,dx &=& 
 \frac{(M_n-M_p)_{QCD}}{m_d-m_u}\simeq 0.41\pm 0.06\, .
 \label{eq:dis13}
\end{eqnarray}

On the other hand, the theoretical expression for the first moment of
$e^{(T=1)} (x)$ is obtained from (\ref{eq:nosing}) as
\begin{equation}
 \int^1_{-1} e^{(T=1)}(x) dx = \frac{N_c}{2I}\frac13
 \sum_a \sum_{n\le 0}\sum_{m>0}\frac{\langle n |\tau_a |m\rangle
 \langle m | \tau_a \gamma^0|n\rangle }{E_m-E_n} \, ,
 \label{eq:modelsum}
\end{equation}
Here, we have used the fact that, since the contribution $e_2(x)$
is a total derivative, it does not contribute
to the integral of (\ref{eq:modelsum}). 
After integration over $x$, the double sum over levels in 
(\ref{eq:modelsum}) is naturally restricted to include only
transitions from occupied to non-occupied states.
This is reasonable, since
the operator appearing in the r.h.s. of (\ref{eq:modelsum})
is a local operator, and transitions from occupied to occupied states
would violate the Pauli principle.
Within the framework of the CQSM, we can evaluate the
r.h.s. of (\ref{eq:dis12}), i.e. the isovector scalar charge of the
nucleon $\langle N|\bar{\psi}\tau^3\psi|N\rangle $, directly
without passing through the distribution function.
Since the resultant expression of $\langle N|\bar{\psi}\tau^3\psi|N
\rangle $ precisely coincides with the r.h.s. of (\ref{eq:modelsum}),
we conclude that the first moment sum rule of $e^{(T=1)}(x)$ is
properly satisfied within the model.

\section{Numerical results and discussion}

The numerical method used for evaluating $e(x)$ in this paper is
essentially the same as the one used for the computing the
twist-2 distributions $q(x),\, \Delta q(x),\,
\delta q(x)$~\cite{WK98,WK99}.
The eigen-energies and the eigen-vectors of the static Dirac
Hamiltonian $H$ with the hedgehog background are obtained by diagonalizing
it with the so-called Kahana-Ripka plane-wave basis~\cite{KR84}.
Following them, the plane-wave states, introduced as a set of
eigenstates of the free Hamiltonian 
$H_0 = - i \mbox{\boldmath $\alpha$}\cdot\! \nabla\! + \beta M$, 
is discretized by imposing an appropriate boundary condition for the 
radial wave functions at the radius $D$ chosen to be sufficiently larger
than the soliton size. The basis is made finite by
retaining only those states with the momentum $k$ satisfying
the condition $k<k_{max}$. As a results of using
this discretized momentum basis, a resultant distribution
becomes a discontinuous function of $x$, due to the factor
$\delta(x M_n-E_n-\hat{p}_3)$.
In order to get a continuous function with a discretized basis,
we introduce a smeared distribution function in the variable $x$
as \cite{DPPPW97}
\begin{equation}
 e_{\gamma}(x)\equiv \frac{1}{\gamma \sqrt{\pi}}
 \int_{-\infty}^{\infty}
 e^{-\frac{(x-x')^2}{\gamma^2}}e(x') dx',
 \label{eq:smeared}
\end{equation}
with a small but finite value of $\gamma$ ($\gamma\ll 1$).
The smeared distribution is expected to be continuous when
the separation between the discretized momenta is much smaller
than the smearing width $\gamma$. Since the physical distribution
corresponds to the limit $\gamma \rightarrow 0$, this 
forces us to employ a very large box size $D$ to get a continuous
distribution function.

This procedure works very well at least for the standard distributions
investigated so far.
However, in the numerical calculation of $e^{(T=0)}(x)$, we
have a new problem which we have not encountered before.
Our expectation is that, if a $\delta(x)$-type singularity really
exists in $e^{(T=0)}(x)$, the corresponding smeared distribution
would have a Gaussian peak centered around $x=0$ with the
width $\gamma$.
The problem here is that the distribution function in question
may also have a piece that is non-singular for all value of $x$.
One might think that the contribution of the singular part can
be disentangled from the total contribution by using the
``unsmearing method" described in \cite{DPPPW97}.
This is not feasible, however, by the following reasons.
First, although the smearing procedure defined with
(\ref{eq:smeared}) preserves the integral value of the
distribution, we have no {\it ad hoc} way to know the overall
coefficient of the $\delta(x)$ term of the
distribution. Secondly, the small $x$ behavior of the nonsingular part of
the distribution would be hard to know, because it is buried
in the very large contribution of smeared $\delta$ function
singularity. This point will be discussed in more detail in the
following subsection.

\subsection{isosinglet distribution $\mbox{\boldmath $e^{(T=0)}(x)$}$}

\ \ \ In the numerical calculation, we fix the pion weak decay constant
$f_{\pi}$ in (\ref{eq:hedgehog})
to its physical value, i.e., $f_{\pi}=93\,$MeV, so that only one
parameter of the model is the dynamical quark mass $M$, which
plays the role of the coupling constant between the pion and the 
effective quark fields. 
Through the present analysis, we use the value of $M=375 \,\mbox{MeV}$,
which is favored from the phenomenology of nucleon low energy
observables.
With $M=375 \,\mbox{MeV}$, we have $\Lambda_1\simeq 627 \,\mbox{MeV}$
and $\Lambda_2\simeq 1589 \,\mbox{MeV}$ from the
conditions (\ref{eq:cond1}), (\ref{eq:cond2}).
The static soliton energy
obtained with these parameters is about $1018 \,\mbox{MeV}$.
We point out that, although the soliton mass emerges about $8 \,\%$
larger than the observed nucleon mass $M_N$, the consistency with
the energy-momentum sum rule of the unpolarized distribution functions
enforces us to use this value for $M_N$ in the following evaluation
of the distribution functions.

We start with showing the numerical equivalence of the final answers
based on the non-occupied representation and the occupied one.
The problem here is the dependence on the cut-off momentum $k_{max}$,
which is introduced to make finite the discretized Kahana-Ripka basis set.
Since the distribution $e^{(T = 0)} (x)$ is ultra-violet finite after 
introduction of the double-subtraction Pauli-Villars regularization, 
one might expect that the answers would be stable as far as one takes 
$k_{max}$ much larger than the second Pauli-Villars cut-off mass 
$\Lambda_2 \simeq 1.6 \,\mbox{GeV}$.
This is not the case, however. As clarified in \cite{WO03}, the
$\delta$-function type singularity in 
$e^{(T = 0)} (x)$ is generated by the contribution of the infinitely 
deep Dirac-sea levels, which are naturally contained in either of the 
three terms, i.e. the main term and the two Pauli-Villars subtraction 
terms. This implies that the singularity, which will appear in the
smeared distribution as a Gaussian peak around $x = 0$ with the width 
$\gamma$, would be reproduced only in the ideal limit of 
$k_{max} \rightarrow \infty$.
To achieve this ideal limit, we therefore use an extrapolation method
explained below. For this extrapolation to be done smoothly, 
we first introduce an energy cut-off into the level
sums~(\ref{eq:e0xunoccu}) and (\ref{eq:e0xoccu}) of the form.
\begin{eqnarray}
 [e^u(x)+e^d(x)] _{non-occupied}^R &=& -N_c M_N
 \sum_{n > 0}  \,
 \langle n \vert \gamma^0  \delta (xM_N - \hat{p}_3 - E_n) \vert n \rangle
 R(E_n) ,
 \label{eq:e0-nonoccu}
 \\ 
 {}[e^u(x)+e^d(x)] _{occupied}^R
 &=& \ N_c M_N 
 \sum_{n \leq 0}  \,
 \langle n \vert \gamma^0  \delta (xM_N - \hat{p}^3 - E_n) \vert n \rangle
 R(E_n) .
 \label{eq:e0-occu}
\end{eqnarray}
Here, $R(E_n)$ is a smooth regulator function with an energy cut-off, 
$E_{max} = \sqrt{k_{max}^2 + M^2}$. For this regulator function, 
we employ here a Gauusian function
\begin{equation}
 R(E_n) = \mbox{exp} \left[-{\left(E_n / E_{max}\right)}^2\right] ,
 \label{eq:reglfunc}
\end{equation}
following Diakonov et al.~\cite{DPPPW97}.
We first compute the level sums (\ref{eq:e0-nonoccu})
and (\ref{eq:e0-occu}) for several values
of $k_{max}$, in the case of masses $M$, $\Lambda_1$,
and $\Lambda_2$ respectively, and then perform 
the Pauli-Villars subtraction, and finally remove the energy cut-off 
by the numerical extrapolation to infinity pointwise in $x$.
In the present study, we use five data (corresponding to 
$k_{max} / M = 12, 16, 20, 24, \mbox{and} \ 28)$ and perform a least
square fit of these data by using a fourth order function of $1/k_{max}$.
\begin{figure}[htb] \centering
\begin{center}
 \includegraphics[width=11.0cm]{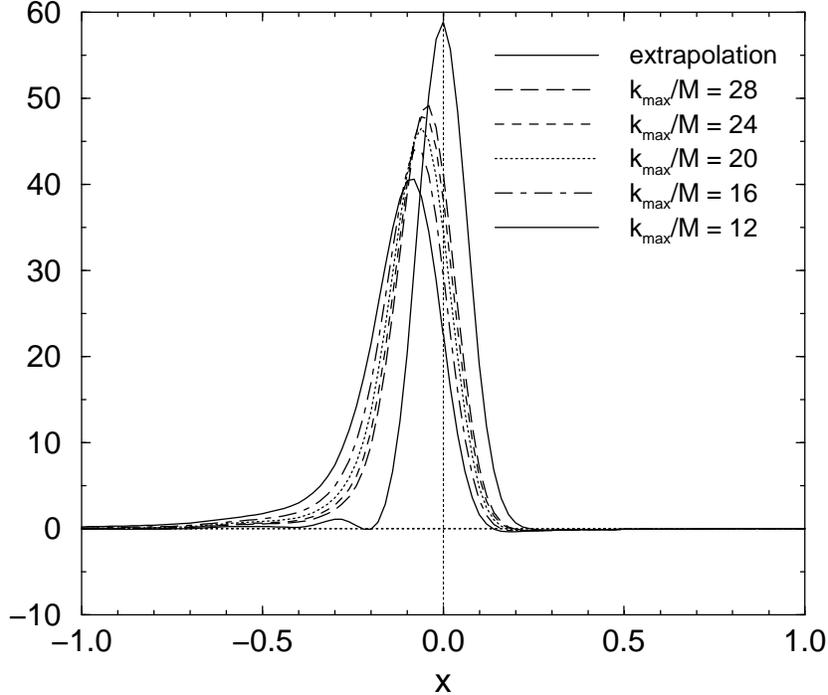}
\end{center}
\vspace*{-0.5cm}
\renewcommand{\baselinestretch}{1.20}
\caption{The $k_{max}$ dependence of the Dirac-sea contribution
to $e^{(T=0)}(x)$ based on the occupied representation.
The solid curve represents the extrapolated result.} 
\label{fig:e0(x).occupied}
\end{figure}

Now we are ready to show in Fig.{\ref{fig:e0(x).occupied}
the $k_{max}$ dependence of the 
Dirac-sea contributions based on the occupied representation for all 
values of $x$. Here we use a value of $\gamma = 0.1$.
This figure shows that the peak positions of the Gaussian-like
function obtained with the finite cutoff energy deviate to the
negative $x$ region from the origin $x = 0$.
This deviation of the peak 
position in the smeared distribution may be understood as follows. 
First, when one uses the occupied representation, the vacuum substraction 
as represented by (\ref{eq:exregl}) is necessary only for the region
$x < 0$, while it is not necessary for $x > 0$, since the vacuum term
identically vanishes for $x > 0$ due to the restriction of the factor
$\delta(x M_N - E_n - \hat{p}_3)$.
Secondly, we recall the fact that the singular term of $e^{(T = 0)} (x)$ 
emerges as a delicate cancellation of two large numbers or the 
infinities, i.e. the difference between the main contribution with 
hedgehog background and the vacuum subtraction term obtained with $U = 1$.
These two facts indicate that the use of the occupied form with
some finite value of $\gamma$ can reproduce 
the redistribution of the delta-function strength at $x = 0$ in the 
$x < 0$ region only, but it cannot do it properly in the $x > 0$ region, 
as far as the finite energy cutoff is used. This is the reason why the 
Gaussian-like peak of the smeared distribution is shifted to the negative 
$x$ region. One can however confirm the behavior that the position of the 
Gaussian peak approaches $x = 0$ as the energy cutoff is 
increased. And, finally, with the extrapolation method, we obtain a
reasonable result which shows that the peak of the smeared distribution
is positioned just at $x= 0$.
(In the above analysis, we fix the box size to
be $DM = 20$. As a matter of course, to get a physically acceptable
answers, we must also investigate the dependence of the answers on
the box size $D$. We found that, above 
$DM = 20$, the change of the small $x$ behavior of $e^{(T = 0)} (x)$ as
illustrated in Fig.{\ref{fig:e0(x).occupied} is almost due to
the increase of $k_{max}$, and the answer is stable against the
further increase of $DM$ above 20.)
\begin{figure}[htb] \centering
\begin{center}
 \includegraphics[width=11.0cm]{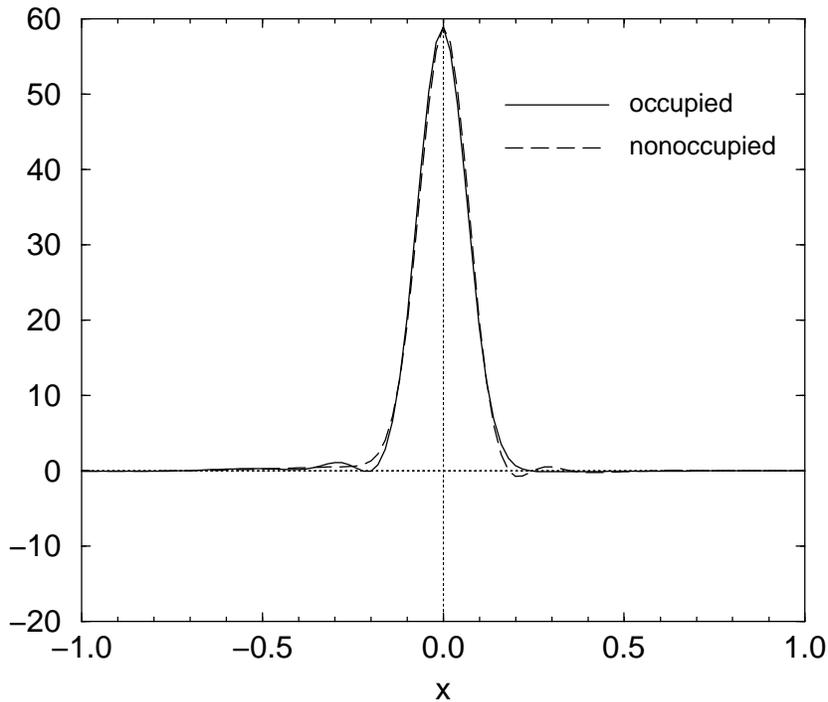}
\end{center}
\vspace*{-0.5cm}
\renewcommand{\baselinestretch}{1.20}
\caption{Comparison of the Dirac-sea contributions to
$e^{(T=0)}(x)$ based on the occupied (solid curve) and
non-occupied (dashed curve) representations.}
\label{fig:occu.and.non-occu}
\end{figure}
After carrying out the similar analysis, this time, with use of the
non-occupied representation, we can now compare the final numerical
results for the Dirac-sea contribution obtained with the two
alternative representations.
Fig.{\ref{fig:occu.and.non-occu} shows this comparison. A reasonable 
agreement between the two ways of evaluating $e^{(T = 0)} (x)$ confirms 
the equivalence of the two representations. At the same time, the
analysis above establishes the existence of the $\delta$-function
singularity in $e^u (x) + e^d (x)$ on the numerical ground.
Some difference between the two curves at the positive and negative $x$
tails of the Gaussian like distributions would be a spurious one
generated by the numerical extrapolation method.
The contributions based on the occupied representation for $x<0$ and
the non-occupied representation for $x>0$ can be obtained after the 
cancellation of two large numbers, i.e., the main contribution and the 
corresponding vacuum subtraction term.
On the other hand, if one uses the occupied representation for $x>0$
and the non-occupied representation for $x<0$, one is free from
the spurious contribution due to the cancellation,
so that the extrapolated curves at these tail region have reasonable
smooth behavior.

Although we were able to confirm the existence of 
$\delta (x)$-type singularity in the numerical analysis of 
$e^{(T = 0)} (x)$, we cannot exclude the possibility that the 
$e^{(T = 0)} (x)$ may also contain a regular term which is smooth in all
the range of $x$. Is it possible to disentangle such non-singular term of 
$e^{(T = 0)} (x)$ from the total contribution containing the singular one?
One should recognize that it is not so easy by the following reasons.
First, the deconvolution method as proposed by Diakonov et al. does not 
work because of the very delicate nature of the
singularity~\cite{DPPPW97}.
Second, we have no {\it ad hoc} way to know the coefficient of
$\delta (x)$ term in the original unsmeared distribution.
Nevertheless, we found that the following trick works for obtaining 
the non-singular distribution excluding the $\delta (x)$ term.
That is, as repeatedly emphasized, by using the non-occupied expression 
for $x < 0$ and the occupied one for $x > 0$, we can avoid the vacuum 
subtraction. Interestingly, this also works to remove the singular
contribution in the bare distribution, and the corresponding smeared
distribution would not contain the Gaussian peak corresponding to
$\delta (x)$-type singularity. (One should remember the fact that the
vacuum term plays an indispensable role in reproducing the
$\delta$-function singularity in $e^{(T=0)}(x)$.)

\begin{figure}[htb] \centering
\begin{center}
 \includegraphics[width=11.0cm]{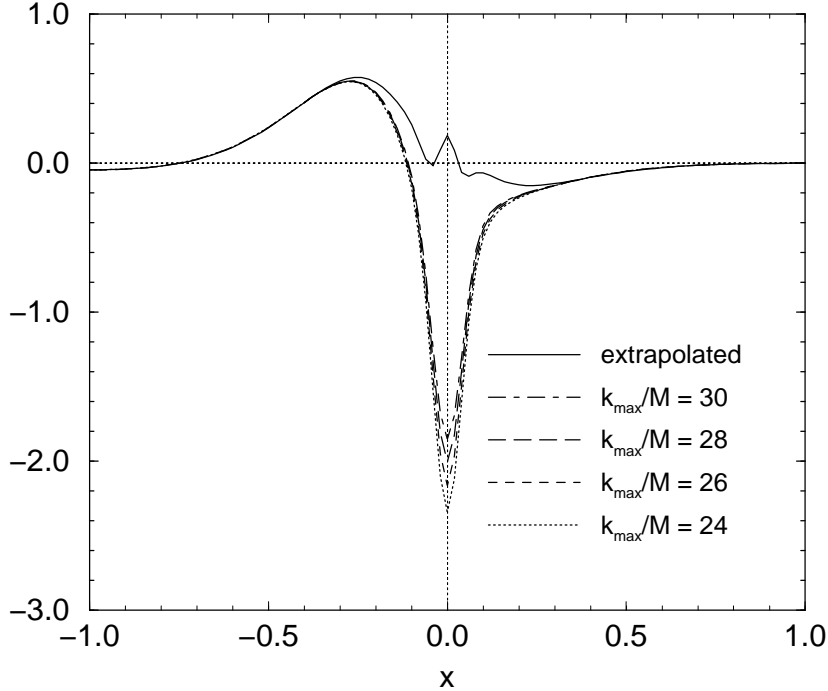}
\end{center}
\vspace*{-0.5cm}
\renewcommand{\baselinestretch}{1.20}
\caption{The $k_{max}$ dependence of  
the Dirac-sea contributions to $e^{(T=0)}(x)$
based on the occupied representation for $x>0$ and the non-occupied 
representation for $x<0$. The solid curves represents
the extrapolated result.}
\label{fig:reg.extrp}
\end{figure}

Fig.~\ref{fig:reg.extrp} shows the $k_{max}$ dependence 
of the Dirac-sea contributions  based on the occupied 
representation for $x>0$ and the non-occupied representation for $x<0$.
One finds that the large and positive Gaussian peak, the reminiscence
of the $\delta$-function singularity in the bare distribution,
does not appear any more.
One can also see that the negative large contributions of the Dirac sea
in the small $x$ region tend to decrease as the cutoff momentum
$k_{max}$ increase.
We again remove the energy cutoff by the numerical extrapolation to
infinity pointwise in $x$. We observe some difference from the
previous case, however. Owing to the feature that the
$\delta$-function singularity is already excluded in the present way
of calculation, the $k_{max}$ dependence is well reproduced by the
linear function of $1 / k_{max}$ as illustrated in
Fig.~\ref{fig:fit.point}.
After this extrapolation procedure, the result
shows a smooth behavior in the whole region of $x$ except the region
$|x|<0.06$ in which the answer is thought to contain some numerical
instability generated by the extrapolation method.  
Neglecting the data in the $|x|<0.06$ region, we make this
extrapolated result smooth.
After deconvoluting the smeared distribution with use of the
Fourier and its inverse transforms, we obtain the final
prediction for the distribution  $e^{(T=0)}(x)$ within the framework
of the CQSM, the normalization point of which may be interpreted as
about $600 \,\mbox{MeV}$.
\begin{figure}[htb] \centering
\begin{center}
 \includegraphics[width=11.0cm]{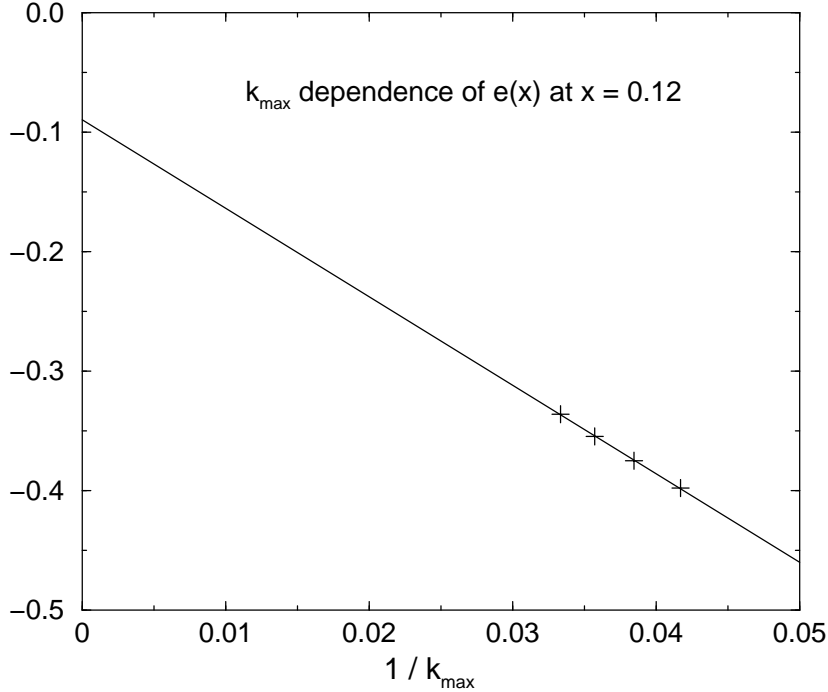}
\end{center}
\vspace*{-0.5cm}
\renewcommand{\baselinestretch}{1.20}
\caption{The $k_{max}$ dependence of $e^{(T=0)}_{sea}(x)$ at
$x=0.12$ and its linear extrapolation to $k_{max} \rightarrow \infty$.}
\label{fig:fit.point}
\end{figure}

Summarizing our analysis up to this point, the isosinglet part of the
chiral-odd twist-3 distribution is given as a sum of the valence quark
and Dirac-sea quark contributions,
\begin{equation}
 e^{(T = 0)} (x) = e^{(T=0)}_{val} (x) + e^{(T=0)}_{sea} (x),
 \label{eq:exvalsea}
\end{equation}
where the Dirac-sea contribution consists of the singular term and the 
nonsingular (regular) term as 
\begin{equation}
 e^{(T=0)}_{sea} (x) = C \delta (x) + e^{(T=0)}_{reg} (x),
 \label{eq:exsea}
\end{equation}
\begin{figure}[htb] \centering
\begin{center}
 \includegraphics[width=11.0cm]{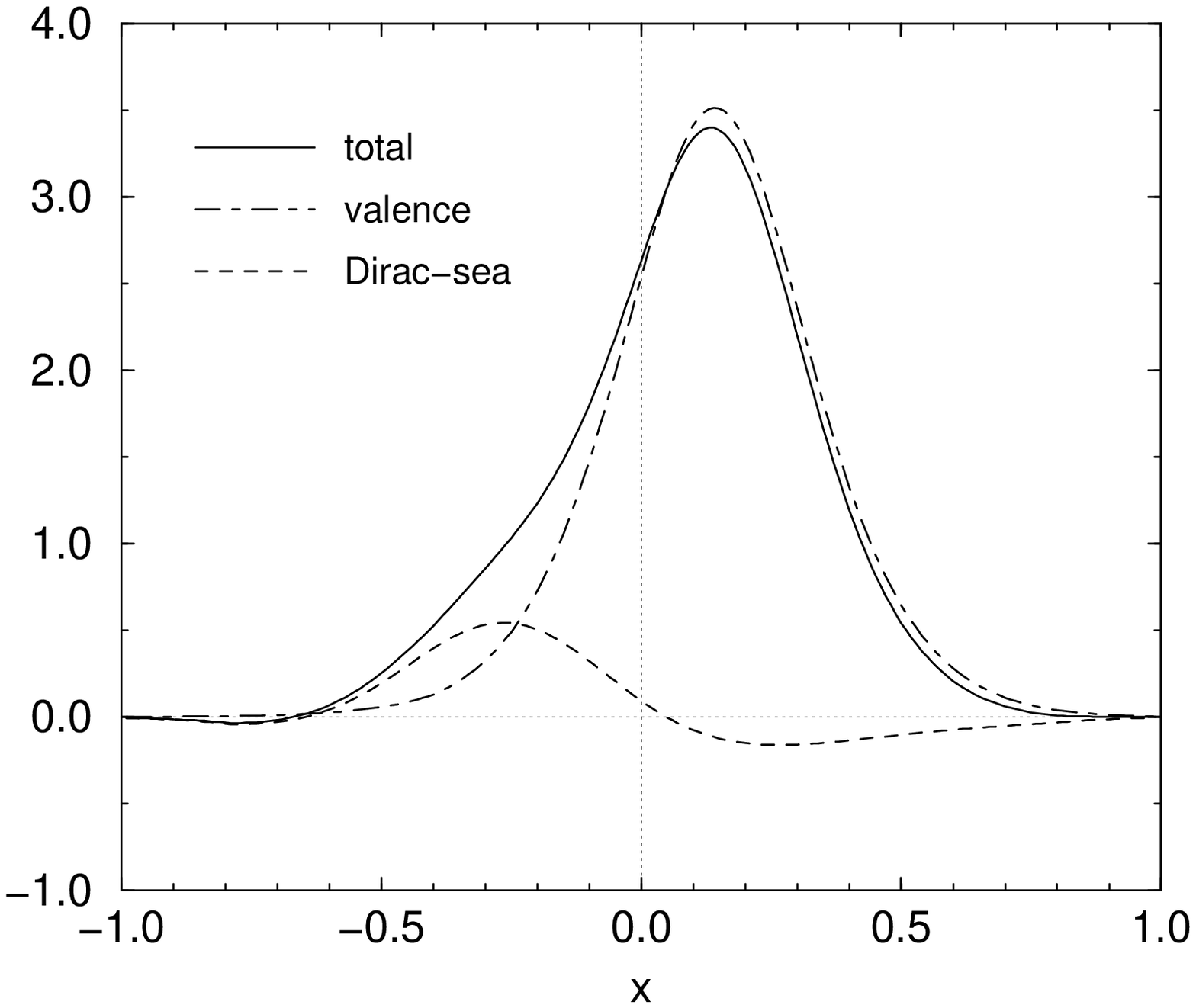}
\end{center}
\vspace*{-0.5cm}
\renewcommand{\baselinestretch}{1.20}
\caption{The final theoretical predictions of the CQSM for $e^{(T=0)}(x)$.
The dot-dashed curve represents the contribution
of $N_c$ valence level quarks, the dashed curve the
non-singular part of the Dirac-sea contributions, and the solid curve
does their sum. The $\delta$-function singularity at $x=0$ in
the Dirac-sea quarks part is not shown in this figure.}
\label{fig:e0(x)}
\end{figure}

Shown in Fig.\ref{fig:e0(x)} are the final theoretical predictions for 
$e^{(T = 0)} (x)$ obtained in the above-explained way.
The dashed curve here represents the contribution of 
$N_c$ valence level quarks, while the dotted curve does the regular part
of Dirac-sea contribution. The sum of these two contributions is shown
by the solid curve.
(We recall that the $\delta (x)$-type singular term is not 
shown in this figure.) One can convince that the regular part of the 
Dirac-sea contribution shows a nontrivial structure in the $x \neq 0$ 
region.

After performing the numerical integration of the above distributions
over $x$, one can obtain the contributions of the valence quark term
and the regular part of the Dirac-sea term to the first moment sum rule :
\begin{eqnarray}
 \int_{-1}^1 e^{(T=0)}_{val} (x) d x &\simeq& 1.7, \\
 \label{eq:exmomval}
 \int_{-1}^1 e^{(T=0)}_{reg} (x) d x &\simeq& 0.18.
 \label{eq:exmomsea}
\end{eqnarray}
Note that the regular part of $e^{(T=0)}_{sea} (x)$ gives small but
non-zero contribution to the sum rule.
To determine the coefficient of the singular 
term in (\ref{eq:exsea}), we use the first moment sum rule
(\ref{eq:ex1stmom-cqsm1}) or (\ref{eq:ex1stmom-cqsm2}) for 
$e^{(T = 0)} (x)$, which was already shown to hold within the framework 
of the CQSM. We first recall that the r.h.s. of the sum rule
(\ref{eq:ex1stmom-cqsm1}) or (\ref{eq:ex1stmom-cqsm2}) is 
the nucleon scalar charge defined by
\begin{equation}
 \bar{\sigma} = \langle N | \bar{\psi}_u \psi_u + \bar{\psi}_d \psi_d 
 | N \rangle .
 \label{eq:scharge}
\end{equation}
The point is that this low energy observable can be calculated within
the CQSM, without asking for the distribution function $e^{(T = 0)} (x)$.
It is given as
\begin{equation}
 \bar{\sigma} = \bar{\sigma}_{val} + \bar{\sigma}_{sea},
 \label{eq:schargevalsea}
\end{equation}
with
\begin{eqnarray}
 \bar{\sigma}_{val} &=& N_c \,\langle 0 | \gamma^0 | 0 \rangle, \\
 \label{eq:schargeval}
 \bar{\sigma}_{sea} &=& N_c \sum_{n < 0} 
 \langle n | \gamma^0 | n \rangle .
 \label{eq:schargesea}
\end{eqnarray}
Numerically, we find that
\begin{equation}
 \bar{\sigma}_{val} \simeq 1.7, \ \ \ 
 \bar{\sigma}_{sea} \simeq 10.1 ,
 \label{eq:scharge.num.valsea}
\end{equation}
so that
\begin{equation}
 \bar{\sigma} = \bar{\sigma}_{val} + \bar{\sigma}_{sea} \simeq 11.8 .
 \label{eq:scharg.num.total}
\end{equation}
Then, by admitting the validity of the first moment sum rule,
one can extract the coefficient of the $\delta (x)$ term as follows :
\begin{equation}
 C = \bar{\sigma}_{sea} - \int_{-1}^1 e^{(T=0)}_{reg} (x) d x 
 \simeq 9.92 .
 \label{eq:coeffc}
\end{equation}
Our procedure for obtaining the coefficient $C$ is different from that
of Schweitzer~\cite{S03}.
After some consideration based on the gradient expansion analysis, he
assumed that the Dirac-sea contribution to $e^{(T = 0)} (x)$ is saturated 
by the $\delta(x)$ term with the coefficient $\Sigma_{\pi N} / m_0$, and 
simply neglected the possible existence of the nonsingular contribution.
In his treatment, then, the nontrivial shape of $e^{(T = 0)} (x)$ at 
$x \neq 0$ solely comes from the contribution of $N_c$ valence level 
quarks. Thus, the total distribution consists of these two terms as
\begin{equation}
 e^{(T = 0)} (x) = \frac{\Sigma_{\pi N}}{m_0} \delta (x) 
 + e_{val} (x) .
 \label{eq:exschweitzer}
\end{equation}
(Here, for simplicity, we ignore the term proportional to the product
of $m_0$ and the unpolarized distribution function.)
In our opinion, this procedure has a 
danger of double counting. Within the framework of the CQSM, the total 
$\pi N$ sigma term divided by the current quark mass $m_0$ is nothing but 
the total scalar charge $\bar{\sigma}$ of the nucleon, which is made up 
of the two terms, $\bar{\sigma}_{val}$ and $\bar{\sigma}_{sea}$.
The $x$-integral of (\ref{eq:exschweitzer}) would then lead to
\begin{equation}
 \int_{-1}^1 e^{(T = 0)} (x) d x 
 = (\bar{\sigma}_{val} + \bar{\sigma}_{sea}) + \bar{\sigma}_{val} ,
 \label{eq:exmomschweitzer}
\end{equation}
which is obviously double counting the valence quark contribution to the 
first moment sum rule. From the practical viewpoint, this double counting 
is not so serious, since $\bar{\sigma}_{val}$ term turns out to be order
of magnitude smaller than $\bar{\sigma}_{sea}$. This dominance of the 
Dirac-sea contribution to the nucleon scalar charge is one of the
distinguishing features of the CQSM. One can say that
it is connected with the unique feature of this model, which is able to
describe simultaneously a localized baryonic excitation 
together with the nontrivial QCD vacuum structure with nonzero quark 
condensate (or nonzero scalar quark density).
In any case, we emphasize that the CQSM predicts fairly large scalar
charge for the nucleon, i.e. $\bar{\sigma} \simeq 11.8$.
Using the current quark 
mass of $m_0 \simeq 5 \,\mbox{MeV}$, as an estimate, this gives 
\begin{equation}
 \Sigma_{\pi N} \equiv m_0 \,\bar{\sigma} \simeq 60 \,\mbox{MeV},
 \label{eq:sigmaterm}
\end{equation}
which seems to favor relatively large values obtained from the 
recent analysis of the pion-nucleon scattering
amplitude~\cite{Koch82}\nocite{PASW99}\nocite{Olsson00}\nocite{PASW02}-\cite{Sainio02}.

Next we turn to the discussion of the second moment sum rule. We first
point out that the $\delta (x)$ term in $e^{(T = 0)} (x)$ does not 
contribute to the second moment. In the CQSM, then, the second moment of 
$e^{(T = 0)} (x)$ receives contributions from two terms in the 
distribution, i.e. the valence quark term $e^{(T = 0)}_{val} (x)$ and the 
regular part of the vacuum polarization term $e^{(T = 0)}_{reg} (x)$. 
After performing the numerical integration, we find that 
\begin{eqnarray}
 \int_{-1}^1 x e^{(T = 0)}_{val} (x) d x &\simeq& 0.23,
 \label{eq:ex2nd1} \\
 \int_{-1}^1 x e^{(T = 0)}_{sea} (x) d x &=& \int_{-1}^1 
 x e^{(T = 0)}_{reg} (x) d x \simeq -0.05 .
 \label{eq:ex2nd2}
\end{eqnarray}
The total second moment is therefore given by 
\begin{equation}
 \int_{-1}^1 x e^{(T = 0)} (x) d x \simeq 0.23 - 0.05 \simeq 0.18 .
 \label{eq:ex2nd3}
\end{equation}
We recall that, within the CQSM, there is another independent method
for evaluating the second moment.
Since we are working in the chiral limit, we rewrite
(\ref{eq:ex2ndmom-cqsm3}), by setting $m_0 = 0$, as
\begin{equation}
 \int_{-1}^1 x e^{(T = 0)}(x) d x = N_c \cdot \frac{M}{M_N} \beta ,
 \label{eq:ex2nd4}
\end{equation}
or
\begin{eqnarray}
 \int_{-1}^1 x e^{(T = 0)}_{val} (x) d x 
 &=& N_c \frac{M}{M_N} \beta_{val}, \\
 \label{eq:ex2nd5}
 \int_{-1}^1 x e^{(T = 0)}_{sea} (x) d x 
 &=& N_c \frac{M}{M_N} \beta_{sea},
 \label{eq:ex2nd6}
\end{eqnarray}
with
\begin{eqnarray}
 \beta_{val} &=& \langle 0 | \frac{1}{2} (U + U^{\dagger}) | 0 \rangle, \\
 \beta_{sea} &=& \sum_{n < 0} \langle n | \frac{1}{2} (U + U^{\dagger})
 | n \rangle .
 \label{eq:ex2nd7}
\end{eqnarray}
These quantities $\beta_{val}$ and $\beta_{sea}$ can be calculated 
directly within the model, without invoking the corresponding 
distribution functions. Numerically, we find that
\begin{eqnarray}
 N_c \frac{M}{M_N} \beta_{val} &\simeq& 0.23, \\
 \label{eq:betval}
 N_c \frac{M}{M_N} \beta_{sea} &\simeq& -0.06.
 \label{eq:betsea}
\end{eqnarray}
These two numbers are consistent with the corresponding numbers in
(\ref{eq:ex2nd1}) and (\ref{eq:ex2nd2}), obtained through the
distribution functions, A small discrepancy between the numbers in
(\ref{eq:ex2nd2}) and (\ref{eq:betsea}) may be interpreted as giving a
measure of numerical errors introduced by the very delicate interpolation
method for obtaining the vacuum polarization term of $e^{(T = 0)} (x)$.
At any rate, we find that the CQSM predicts relatively small but 
nonzero value for the second moment of $e^{(T = 0)} (x)$. Since we are 
working in the chiral limit ($m_0 = 0$), this appears to contradict the 
corresponding sum rule (\ref{eq:ex2ndmom}) derived on the basis of
the QCD equations of motion, which states that the second moment
of $e^{(T = 0)} (x)$ vanishes in the chiral limit.
Does this discrepancy simply mean the limitation of the CQSM as an 
effective theory of QCD?
In our opinion, this is not necessarily the case by the following reasons.
First of all, we point out that moment sum rules containing quark masses 
are somewhat delicate one, since the masses are generally dependent on
the renormalization scale. Secondly, if the QCD vacuum breaks the chiral
symmetry spontaneously as is generally 
believed, a quark acquires a dynamical mass of several hundred MeV, which 
means that massless quarks are nowhere. Naturally, the situation is not
so simple because of the color confinement.
For instance, according to the picture of the MIT bag model, which realizes quark confinement by hand, at least the vacuum inside the bag is 
perturbative and the quarks inside it remains massless.
According to Shuryak~\cite{Shuryak88},
the bag model is based on the idea that the hadron is a piece of 
a qualitatively different (or ``perturbative") phase of the QCD vacuum.
The physical picture of the CQSM for the baryon and the QCD vacuum is 
fairly different from that of the bag model. According to the words of
Shuryak again, the chiral models (including the CQSM) assume that the 
vacuum is only slightly modified inside the hadron : the relative
orientation of the right- and the left-handed quark fields is somewhat
different. This last statement denotes the fact 
that, in the basic lagrangian of the CQSM, the dynamical quark mass 
parameter $M$ appears as a product with the chiral field
$U^{\gamma_5} (x)$, which is space-time dependent.
It is also the cause of the fact that the product of $M$ and $\beta$
enters the r.h.s. of the second moment sum
rule (\ref{eq:ex2ndmom-cqsm3}).
This supports Schweitzer's viewpoint~\cite{S03} that the quantity
$\beta M$ can be interpreted as an effective mass of quarks bound in
the soliton background at least in the 2nd moment sum rule of
$e^{(T=0)}(x)$. Numerically, we have
\begin{equation}
 \beta M \sim 51 \,\mbox{MeV} .
 \label{eq:effmass}
\end{equation}
This value is smaller than the one obtained in \cite{S03}, since the
contribution of the Dirac-sea quarks neglected in \cite{S03} works
to reduce the value of $\beta$.

In any case, the nonzero value of the second moment of $e^{(T = 0)} (x)$
is nothing contradictory at least within the framework of the CQSM
in which massless quarks are nowhere at the model energy scale of
about $600 \,\mbox{MeV}$.
However, we anticipate that the dynamical quark mass $M$ is generally a
scale dependent quantity which approaches zero in the high energy limit.
The naive QCD sum rule for the second moment of $e^{(T = 0)} (x)$
would be recovered in this limit.
To verify the validity of this idea, what is crucial is experimental 
determination of the second moment sum rule at the relatively low energy 
scale close to the above-mentioned model energy scale.
This may be in principle possible by inversely evoluting high energy
data to low energy scale.

\subsection{isovector distribution $\mbox{\boldmath $e^{(T=1)}(x)$}$}

In the case of isovector distribution $e^{(T=1)}(x)$,
no ultraviolet regularization is needed because its first moment,
(\ref{eq:modelsum}), is related to the imaginary part of the
Euclidian effective meson action in the background
soliton field~\cite{GPPSU01} and it is ultraviolet finite.
We have checked that the energy level sum (\ref{eq:nosing}) is
stable enough against the increase of the cutoff momentum $k_{max}$,
above $12M$. The final result for the isovector distribution
$e^{(T=1)}(x)$ is shown in Fig.~\ref{fig:e1(x)}.
\begin{figure}[htb] \centering
\begin{center}
 \includegraphics[width=11.0cm]{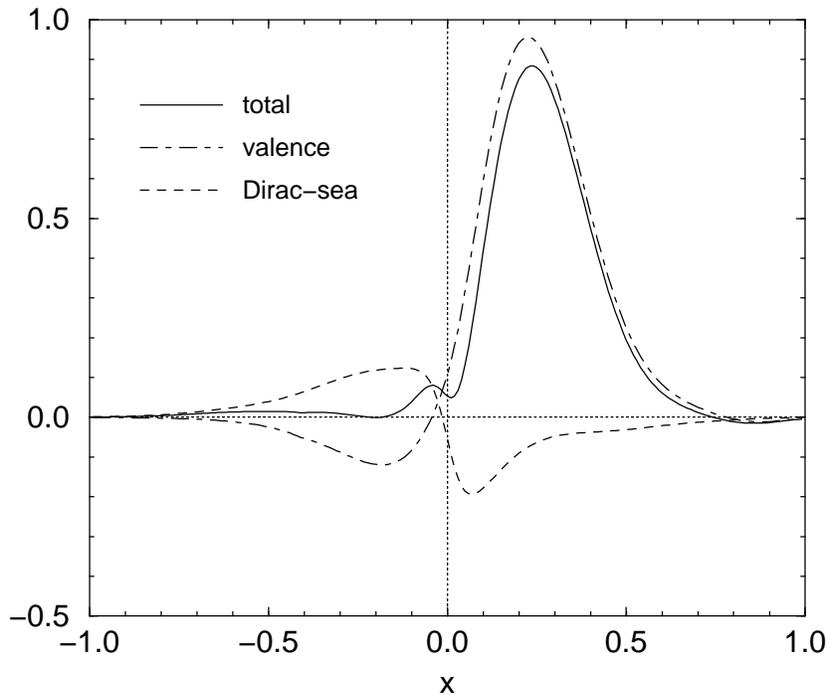}
\end{center}
\vspace*{-0.5cm}
\renewcommand{\baselinestretch}{1.20}
\caption{The theoretical predictions of the CQSM for $e^{(T=1)}(x)$.
The dot-dashed curve stands for the contribution
of $N_c$ valence level quarks, the dashed curve the contribution
of the Dirac-sea quarks, while the solid curves represents
their sum.}
\label{fig:e1(x)}
\end{figure}%
The dashed curve represents the contribution of the $N_c$ valence level
quarks, the dot-dashed curve represents the contribution of
the Dirac-sea quarks, while the solid curve represents their sum.
In contrast to the isosinglet distribution, the Dirac-sea contribution
has no singularity at $x=0$ and it is a smooth function
in the whole region of $x$.
The total contribution is given by the solid curve.

\begin{figure}[htb] \centering
\begin{center}
 \includegraphics[width=11.0cm]{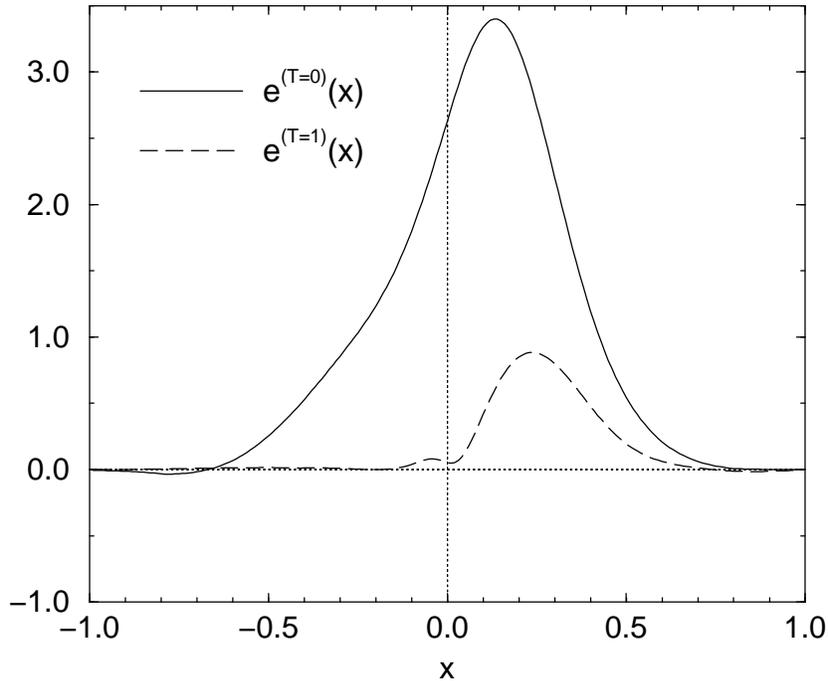}
\end{center}
\vspace*{-0.5cm}
\renewcommand{\baselinestretch}{1.20}
\caption{The comparison of the theoretical predictions for
$e^{(T=0)}(x)$ and $e^{(T=1)}(x)$ at the model energy scale.}
\label{fig:e0.vs.e1}
\end{figure}%
The first moment or the $x$-integral of this total contribution 
gives the following value
\begin{equation}
 \int_{-1}^1 e^{(T = 1)} (x) d x \simeq 0.28 ,
\end{equation}
which is order of magnitude consistent with the estimate obtained from 
the analysis of the non-electromagnetic proton-neutron mass difference.
Shown in Fig.\ref{fig:e0.vs.e1} are the comparison of our final
theoretical predictions for $e^{(T = 0)} (x)$ and $e^{(T = 1)} (x)$.
One confirms that the magnitude of $e^{(T = 1)} (x)$ is much smaller
than that of $e^{(T = 0)} (x)$ in conformity 
with the large $N_c$ relation (\ref{eq:N_c-rel}).
Combining these two distributions, we can now give final theoretical
predictions for the chiral-odd twist-3 distribution function $e^a (x)$
of each flavor $a$.
Shown in Fig.\ref{fig:eu.ed}(a) are the distributions for the $u$-quark
and the $\bar{u}$-quark, while Fig.\ref{fig:eu.ed}(b) gives the
distributions for the $d$-quark and the $\bar{d}$-quark.

\begin{figure}[htb] \centering
\begin{center}
 \includegraphics[width=15.0cm]{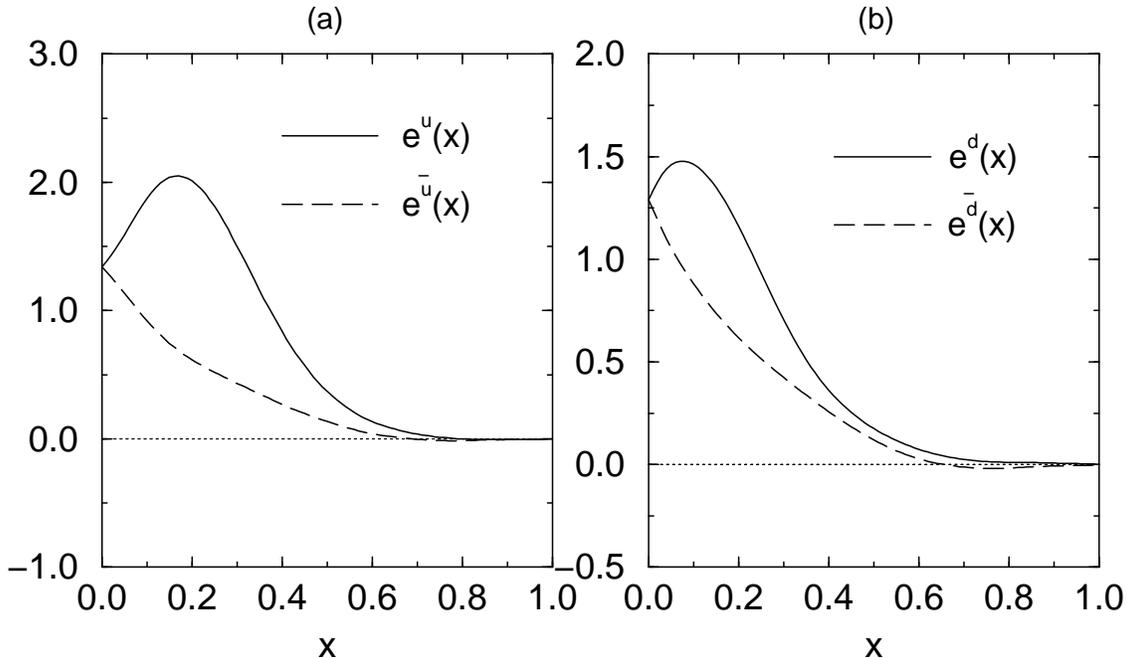}
\end{center}
\vspace*{-0.5cm}
\renewcommand{\baselinestretch}{1.20}
\caption{The theoretical predictions for $e^u(x)$, $e^{d}(x)$,
$e^{\bar{u}}(x)$, and $e^{\bar{d}}(x)$ at the model energy scale.}
\label{fig:eu.ed}
\end{figure}%

\subsection{comparison with the empirical information from the CLAS
measurements}

Here, we make a very preliminary comparison of our theoretical 
predictions for $e(x)$ with the empirical information extracted from the
high-energy semi-inclusive scatterings.
Because of its chiral-odd nature, the distribution function $e(x)$
does not appear in inclusive DIS cross sections.
To extract any information for it, we must therefore carry out 
more specific semi-inclusive type scattering experiments.
Very recently, such an experiment has in fact been done by the CLAS
Collaboration~\cite{CLAS}.
They measured the azimuthal asymmetry $A_{LU}$ in the
electro-production of pions from deeply inelastic scatterings of
longitudinally polarized electrons off unpolarized protons. 

The first theoretical analysis of the CLAS data was carried out by
Efremov et al.~\cite{EGS01},\cite{EGS02}.
Their analysis assume that the beam single spin
asymmetry measured by the CLAS group is dominantly generated by
the so-called Collins mechanism~\cite{Collins93}.
Under this assumption together with a particular
parameterization for the Collins fragmentation function, they were
able to extract the first information on the chiral-odd twist-3
distribution function $e(x)$.
Recently, this analysis was criticized by Yuan~\cite{Yuang03}.
According to him, there may be another mechanism which competes
with the Collins mechanism~\cite{Sivers90},\cite{Sivers91}.
It is the leading order transverse momentum dependent parton
distribution $h^{\bot}_1 (x, k_{\bot})$ convoluted with chiral-odd
fragmentation function $\hat{e} (z)$.
After all, the fact is that we still have poor knowledge about the
mechanism that generates the beam single spin asymmetry in
semi-inclusive deep inelastic scatterings. We must understand the
mechanism of parton fragmentation processes into hadrons, especially
the physics of time-reversal-odd fragmentation
functions~\cite{Collins93},\cite{MT96}.
We must also clarify the dynamics of transverse momentum dependent
parton distribution functions in combination with the physics of
chiral-odd fragmentation
functions~\cite{Sivers90}\nocite{Sivers91}-\cite{MT96}.
A truly reliable extraction of the chiral-odd twist-3
distribution function $e(x)$,
which is of our primary concern here, can be achieved only after
more complete understanding of the above-mentioned mechanisms of
semi-inclusive DIS processes.

Keeping this fact in mind, we shall proceed here by assuming the 
dominance of the Collins mechanism. Under this assumption, the
asymmetry measured by the CLAS experiment is interpreted to be
proportional to 
\begin{equation}
 A^{\sin \phi}_{LU} \sim -\frac{4 \pi \alpha^2 s}{Q^4} \,\lambda_e \,
 2 y \sqrt{1 - y} \sum_a e^2_a x^2 e^a (x) H^{\bot a}_1 (z) ,
\end{equation}
with $y = (P \cdot q) / (P \cdot l), z = (P \cdot p_h) / (P \cdot q)$
and $s$ is the invariant mass squared of the photon-hadron system
in the notation of \cite{EGS01}. $\lambda_e$ denotes the beam helicity.
The chiral- and T-odd twist-2 ``Collins" fragmentation function
$H^{\bot a}_1 (z)$ gives the probability of a spinless or unpolarized
hadron to be created from a transversely polarized scattered quark.
Using informations on $H^{\bot a}_1 (z)$ from HERMES
data~\cite{herm01},\cite{herm02},
one can then get direct information on the distribution function
$e(x)$~\cite{EGS01},\cite{EGS02}.
In the CLAS experiment, the azimuthal asymmetries
$A^{\sin \phi}_{LU}$ for the process
$\vec{e} p \rightarrow e^\prime \pi^+ X$ were measured at
$Q^2 \sim 1.5 \,\mbox{GeV}^2$.
Under the dominant-flavor-only approximation for the fragmentation
functions, the semi-inclusive $\pi^+$ production measures the
following combination of the distributions,
\begin{equation}
 e^u (x) + \frac{1}{4} e^{\bar{d}} (x) .
\end{equation}
In Fig.~\ref{fig:with-clas}, we make a comparison between the
predictions of the CQSM for the above combinations of the distributions
and the corresponding empirical information
extracted from the CLAS data by Efremov et al.~\cite{EGS01},\cite{EGS02}
under the assumption of Collins mechanism dominance.
The theoretical distribution here corresponds to the energy scale of
$Q^2 = 1.5 \,\mbox{GeV}^2$.
The scale dependence of the distribution is taken
into account by solving the leading-order DGLAP type equation
obtained in the large $N_c$ limit~\cite{KN97}.
(The starting energy scale of
this evolution is taken to be $Q^2_{ini} \simeq 0.30 \,\mbox{GeV}^2$.)
The distribution $e^u (x) + \frac{1}{4} e^{\bar{d}} (x)$ extracted
from the CLAS data contains large errors mainly due to the large
uncertainties of $H^{\bot}_1 (z)$ from the HERMES
data~\cite{herm01},\cite{herm02}.
Still, it was emphasized in \cite{EGS01},\cite{EGS02} that
the extracted distribution
is definitely larger than the ``twist-3 bound" and about two times
smaller than the corresponding unpolarized distribution $f^a_1 (x)$
at the same energy scale. One sees that our theoretical prediction for
$e^u (x) + \frac{1}{4} e^{\bar{d}} (x)$ is in fairly good agreement
with the extracted behavior from the CLAS data. The relatively small
magnitude of the extracted $e(x)$ indicates that there must be a
significant contribution to the $\pi N$ sigma-term sum rule from
the small $x$ region. Whether this is due to the indicated
$\delta$-function singularity in $e(x)$ or it is due to the
yet-unresolved Regge behavior in the small $x$ region is difficult
to judge at the present stage of study.
It is highly desirable to extend the region of measurements to
smaller $x$ region. This is important, because the unambiguous
establishment of the violation of the $\pi N$ sigma-term sum rule
would indirectly prove the existence of a novel $\delta$-function
singularity in the distribution function $e(x)$ of the nucleon,
which in turn may be interpteted as a manifestation of the nontrivial
structure of QCD vacuum in an observable of a localized QCD excitation,
i.e. the nucleon.

\begin{figure}[htb] \centering
\begin{center}
 \includegraphics[width=11.0cm]{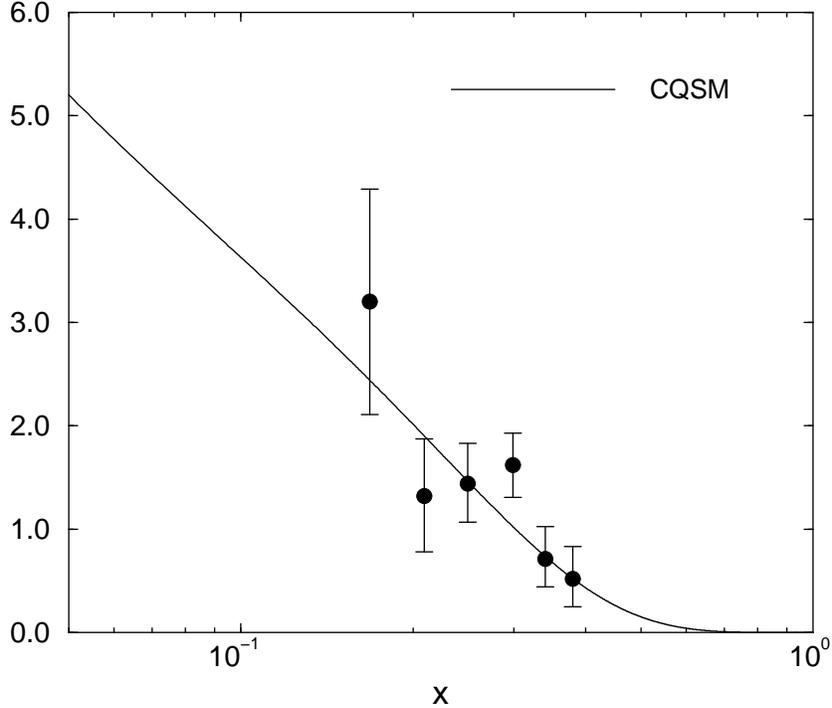}
\end{center}
\vspace*{-0.5cm}
\renewcommand{\baselinestretch}{1.20}
\caption{The theoretical prediction for
$e(x)=e^u(x)+\frac14e^{\bar{d}}(x)$ in comparison with the
corresponding empirical information extracted from the CLAS data at
$\langle Q^2\rangle =1.5\mbox{GeV}^2$ under the assumption of
Collins mechanism dominance.}
\label{fig:with-clas}
\end{figure}

Finally, we want to make some comments on the prediction for $e(x)$ 
based on the MIT bag model. As mentioned in \cite{EGS02}, the bag model
prediction of \cite{Signal97} evolved to the comparable energy
scale of $Q^2 = 1 \,\mbox{GeV}^2$ is in qualitative agreement with the
extracted $e(x)$ from the CLAS data in \cite{EGS02}.
In our opinion, this agreement should be
taken as fortuitous by the following reason. First, as already pointed
out, the isosinglet scalar charge of the nucleon predicted by the
MIT bag model is only about $15 \%$ of the value expected from the
phenomenological knowledge of the $\pi N$ sigma-term.
The fact is that the nucleon isoscalar charge is a quantity of 
order 1 (or order $N_c$, more precisely) in the MIT bag model
or in any other models which contains
three valence-quark degrees of freedom only. 
The situation is totally different in the CQSM. Although the
contribution of the $N_c$ valance level quarks is of the same order
as that of the MIT bag model, the vacuum polarization effect or the
contribution of the Dirac-sea quarks give nearly seven times larger
contribution as compared with that of the valence quarks, thereby
reproducing the correct magnitude of the nucleon scalar charge or
the $\pi N$ sigma-term. Unfortunately, this crucial difference
between the two models is not reflected in the observable distribution
function $e(x)$. Since the Dirac-sea contribution in the CQSM is nearly
saturated by the $\delta$-function singularity, it happens that the
distributions $e(x)$ at $x \neq 0$ predicted by the two models are not
extremely different with each other. This is the reason why the
naive MIT bag model, which fails to explain the magnitude of the
$\pi N$ sigma-term, can reproduce the empirical distribution $e(x)$
extracted from the CLAS data at least qualitatively.

\begin{figure}[htb] \centering
\begin{center}
 \includegraphics[width=11.0cm]{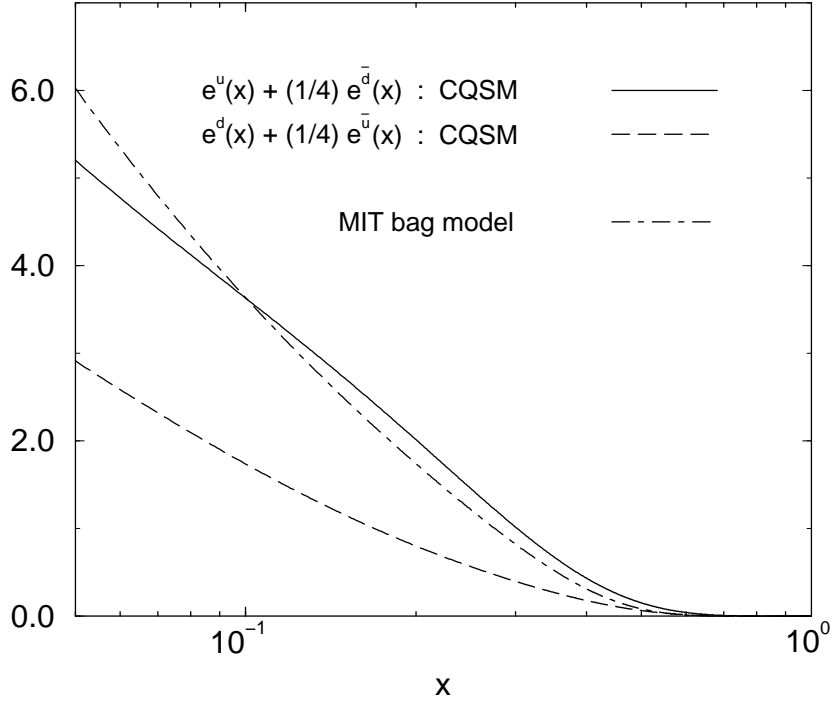}
\end{center}
\vspace*{-0.5cm}
\renewcommand{\baselinestretch}{1.20}
\caption{The predictions of the CQSM for $e^u(x)+(1/4)e^{\bar{d}}(x)$
and $e^d(x)+(1/4)e^{\bar{u}}(x)$ evolved to the energy scale
$Q^2 \simeq 1.5 \,\mbox{GeV}^2$ of the CLAS data from the initial scale
of the model $Q^2_{ini} \simeq 0.30 \,\mbox{GeV}^2$. Also shown for
comparison is the prediction of the MIT bag model evolved to the
same scale from somewhat lower energy scale of $Q^2_{ini} \simeq
0.16 \,\mbox{GeV}^2$.}
\label{fig:pi+.pi-}
\end{figure}

Still, we will show that there are some qualitative and observable
differences between the predictions of the CQSM and the MIT bag model.
The key observation here is that, for the spin-independent chiral-odd  
twist-3 distribution functions, the MIT bag model predicts no flavor 
dependence. That is, within the framework of the naive MIT bag model, 
we have
\begin{equation}
 e^u (x) = e^d (x), \ \ 
 e^{\bar{u}} (x) = e^{\bar{d}} (x) ,
\end{equation}
or more specifically
\begin{equation} 
 e^u (x) + \frac{1}{4} e^{\bar{d}} (x) 
 = e^d (x) + \frac{1}{4} e^{\bar{u}} (x) .
\end{equation}
Such equalities can be expected to hold only in the fictitious limit
of $N_c \rightarrow \infty$. As is in fact the case with the CQSM,
for a finite value of $N_c$, the isovector distribution
$e^{(T = 1)} (x) = e^u (x) - e^d (x)$ does not vanish, so that we
definitely expect that
\begin{equation}
 e^u (x) + \frac{1}{4} e^{\bar{d}} (x) 
 \neq e^d (x) + \frac{1}{4} e^{\bar{u}} (x) .
\end{equation}
Fig.~\ref{fig:pi+.pi-} shows the comparison of the predictions of
the two models for the distributions
$e^u (x) + \frac{1}{4} e^{\bar{d}} (x)$ and
$e^d (x) + \frac{1}{4} e^{\bar{u}} (x)$ evolved to the energy scale
of CLAS experiment, i.e. $Q^2 \simeq 1.5 \,\mbox{GeV}^2$ from the
initial energy scale of the model $Q^2_{ini} \simeq 0.30 \,\mbox{GeV}^2$.
The solid and dashed curves here stand for the predictions of the
CQSM, respectively for $e^u (x) + \frac{1}{4} e^{\bar{d}} (x)$ and
$e^d (x) + \frac{1}{4} e^{\bar{u}} (x)$. On the other hand,
the dot-dashed curve represents the prediction of the MIT bag model,
which gives an identical answer for both these combinations of
distributions. One sees that the CQSM predicts a sizably large
difference between the two distributions
$e^u (x) + \frac{1}{4} e^{\bar{d}} (x)$ and 
$e^d (x) + \frac{1}{4} e^{\bar{u}} (x)$, in sharp contrast to
the MIT bag model.
In principle, the possible differences of these two distributions
can be detected by performing a comparative analysis of the
semi-inclusive $\pi^+$ and $\pi^-$ productions.

\section{Summary and Conclusion}

In summary, we have given theoretical predictions for the chiral-odd 
twist-3 distribution function $e^a (x)$ of the nucleon with each flavor
$a$ on the basis of the chiral quark soliton model.
A prominent feature of the isosinglet combination of the distributions,
$e^u (x) + e^d (x)$, is that its first moment is proportional to the
familiar $\pi N$ sigma-term and that it 
contains a delta-function singularity at $x = 0$. In the previous study 
based on the derivative expansion technique, we demonstrated that the
physical origin of this singularity can be traced back to the long-range
quark-quark correlation of scalar type, which signals the
spontaneous chiral symmetry breaking of the QCD vacuum.
The present calculation, without recourse to 
the derivative expansion type approximation, has revealed the following 
facts. The isosinglet distribution $e^u (x) + e^d (x)$ consists of two 
parts, i.e. the contribution of $N_c$ valence level quarks and that of 
the Dirac sea quarks in the hedgehog mean field.
The former takes a familiar shape of distribution 
which has a peak around the value of $x \simeq 1/3$. On the other hand, 
the latter certainly contains a $\delta$-function
type singularity at $x = 0$, but it 
also has nontrivial support for $x \neq 0$. The isovector 
distribution $e^u (x) - e^d (x)$ also consists of the valence and 
Dirac-sea contributions. For this distribution, however, no delta-function 
type singularity is observed, which means that it is a regular function in
all the range of $x$.

The moment sum rules of $e(x)$ provide us with 
valuable information concerning the basic dynamical content of the model
in view of the underlying theory, i.e. QCD.
We showed that the first moment sum 
rule for $e^u (x) + e^d (x)$ is satisfied within the model, if and only
if the delta-function singularity is properly taken into account.
Note however that the delta-function term alone does not saturate the
first moment or the $\pi N$ sigma-term sum rule in contrast to the 
previous argument based on the framework of the perturbative QCD.
We also pointed out that the
second moment sum rule for $e^u (x) + e^d (x)$ does not vanish even in
the chiral limit in contrast to the QCD-equation-of-motion argument.
In our opinion, this violation of the second moment sum rule does not
necessarily show a defect of the model.  It is rather to be 
interpreted as showing the limitation of the perturbative analysis
as a tool of handling a bound state problem and/or the problem of
masses nonperturbatively generated by the mechanism of the
spontaneous chiral-symmetry breaking.
We have also shown that the model prediction for the first moment of the
isovector distribution $e^u (x) - e^d (x)$ comes out to be order of 
magnitude consistent with the phenomenological estimate obtained from
the nonelectromagnetic neutron-proton mass difference.

It was shown that the theoretical predictions for the
distribution $e^u (x) + \frac{1}{4} e^{\bar{d}} (x)$ are in a good 
agreement with the corresponding empirical information
extracted from the CLAS data for the semi-inclusive $\pi^+$
production under the assumption of the Collins mechanism dominance.
This agreement, combined with our analysis explained in the text,
impiles the existence of $\delta$-function singularity at $x = 0$
in the isosinglet distribution $e^u (x) + e^d (x)$,
although a definite conclusion must awaits for
more complete measurements and more thorough 
understanding of the reaction mechanism that generates the beam single 
spin asymmetry in the semi-inclusive pion productions.

 Finally, we compare our theoretical predictions with those of the MIT 
bag model. As shown in the body of the paper, the two models give
accidentally close predictions for the distribution function 
$e^u (x) + \frac{1}{4} e^{\bar{d}} (x)$ at $x \neq 0$. We have shown,
however, that the CQSM predicts a sizably large difference between the 
two distributions $e^u (x) + \frac{1}{4} e^{\bar{d}} (x)$ and $e^d (x) 
+ \frac{1}{4} e^{\bar{u}} (x)$, for which the MIT bag model makes no 
difference. The predicted sizable difference between 
the two combinations of distributions will be detected by performing 
a comparative experimental analysis of the semi-inclusive 
$\pi^+$ and $\pi^-$ productions.

\newpage
\bibliographystyle{unsrt}
\bibliography{ex3}

\end{document}